\documentclass[final,twoside,11pt]{entics} 
\usepackage{graphicx}
\usepackage[all]{xy}
\usepackage{ifthen}
\usepackage{xspace}

\newboolean{talk}
\setboolean{talk}{false}
\newboolean{paper}
\setboolean{paper}{false}

\newboolean{IEEEstyle}
\setboolean{IEEEstyle}{false}
\newboolean{lipicsstyle}
\setboolean{lipicsstyle}{false}
\newboolean{eptcsstyle}
\setboolean{eptcsstyle}{false}
\newboolean{enticsstyle}
\setboolean{enticsstyle}{false}
\newboolean{sigplanstyle}
\setboolean{sigplanstyle}{false}
\newboolean{easychairstyle}
\setboolean{easychairstyle}{false}
\newboolean{scrartclstyle}
\setboolean{scrartclstyle}{false}
\newboolean{lncsstyle}
\setboolean{lncsstyle}{false}

\newboolean{acmartstyle}
\setboolean{acmartstyle}{false}
\newboolean{mfpsstyle}
\setboolean{mfpsstyle}{false}
\newboolean{needstheorems}
\setboolean{needstheorems}{false}
\newboolean{withimages}
\setboolean{withimages}{false}
\newboolean{withproofs}
\setboolean{withproofs}{true}

\newboolean{french}
\setboolean{french}{false}

\usepackage{tikz}
\usetikzlibrary{calc}
\usetikzlibrary{matrix,arrows}
\usetikzlibrary{decorations.shapes}
\usetikzlibrary{decorations.text}
\usetikzlibrary{decorations.pathmorphing}
\usetikzlibrary{decorations.markings}
\usetikzlibrary{positioning}

\tikzset{
node distance=1.3cm, auto,
every node/.style={font=\scriptsize },
ocenter/.style={baseline={([yshift=-.5ex, xshift=-.5ex]current bounding box)}},  
labelBeginAbove/.style={postaction={decorate,decoration={markings,mark=at position 0 with {\node[inner sep= 0.6pt, above=1pt]{\tiny #1};}} } },
labelBeginBelow/.style={postaction={decorate,decoration={markings,mark=at position 0 with {\node[inner sep= 0.6pt, below=1pt]{\tiny #1};}}}},
labelEndAbove/.style={postaction={decorate,decoration={markings,mark=at position 1 with {\node[inner sep= 0.6pt, above=1pt]{\tiny #1};}}}},
labelEndBelow/.style={postaction={decorate,decoration={markings,mark=at position 1 with {\node[inner sep= 0.6pt, below=1pt]{\tiny #1};}}}},
labelEndRight/.style={postaction={decorate,decoration={markings,mark=at position 1 with {\node[inner sep= 0.6pt, right=1pt]{\tiny #1};}}}},
labelEndLeft/.style={postaction={decorate,decoration={markings,mark=at position 1 with {\node[inner sep= 0.6pt, left=1pt]{\tiny #1};}}}}
}

\newcommand{\med}[2]{
($(#1)!.5!(#2)$)
}
\setboolean{enticsstyle}{true}

\usepackage{amsfonts,amssymb,amsmath,mathtools}
\usepackage[T1]{fontenc}
\usepackage{graphicx}
\usepackage{hyperref}
\usepackage{color}

\usepackage{booktabs}   %% For formal tables:
\usepackage{subcaption} %% For complex figures with subfigures/subcaptions
\usepackage[all]{xy}
\usepackage{booktabs}   %% For formal tables:
\usepackage{subcaption} %% For complex figures with subfigures/subcaptions
\usepackage[normalem]{ulem}
\usepackage{multirow}

\usepackage[utf8]{inputenc}

\usepackage{multirow}
\usepackage{cmll}
\usepackage{listings}
\usepackage{stmaryrd}
\usepackage{mathtools}
\usepackage{listings}

\usepackage{fancybox}
\usepackage{hhline}
\usepackage{marginnote}
\usepackage{soul}
\usepackage{xcolor}

\usepackage{cleveref}
\usepackage{complexity}

\newcommand{\semb}[1]{\blue{\llbracket}#1\blue{\rrbracket}}

\newcommand{\ignore}[1]{}
\newcommand{\colspace}{@{\hspace{.3cm}}}
\newcommand{\myinput}[1]{\ifthenelse{\boolean{withimages}}{\input{#1}}{}}

\newcommand{\reflemma}[1]{Lemma~\ref{l:#1}}
\newcommand{\refthm}[1]{Thm.~\ref{thm:#1}}
\newcommand{\refprop}[1]{Prop.~\ref{prop:#1}}

\newcommand{\refsect}[1]{Sect.~\ref{sect:#1}}

\newcommand{\refapp}[1]{Appendix~\ref{sect:#1} (p.~\pageref{sect:#1})}

\renewcommand{\refeq}[1]{(\ref{eq:#1})} %renewcommand to avoid conflict with package mathtools
\newcommand{\reffig}[1]{Fig.~\ref{fig:#1}}

\newcommand{\ie}{\textit{i.e.}\xspace}
\newcommand{\eg}{\textit{e.g.}\xspace}

\newcommand{\ES}{\text{ES}\xspace}
\newcommand{\ESs}{\text{ESs}\xspace}

\newcommand{\blue}[1]{{\color{blue} {#1}}}

\newcommand{\defeq}{\coloneqq} %Giulio
\newcommand{\grameq}{\Coloneqq} %Giulio
\newcommand{\set}[1]{\{#1\}}

\newcommand{\size}[1]{|#1|}

\newcommand{\gc}{{\sf gc}}

\renewcommand{\l}{\lambda}
\newcommand{\isub}[2]{\{#1/#2\}}
\newcommand{\replace}[2]{#1{\shortleftarrow}#2}
\renewcommand{\isub}[2]{\{\replace{#1}{#2}\}}
\newcommand{\esub}[2]{[\replace{#1}{#2}]}
\renewcommand{\esub}[2]{[#1{\shortleftarrow}#2]}

\newcommand{\fv}[1]{{\sf fv}(#1)}

\newcommand{\rootRew}[1]{\mapsto_{#1}}

\newcommand{\Rew}[1]{\rightarrow_{#1}}
\newcommand{\Rewn}[1]{\rightarrow_{#1}^*}

\newcommand{\lRew}[1]{\; \mbox{}_{#1}{\leftarrow}\ }

\newcommand{\lto}{\lRew{}}

\newcommand{\rtom}{\rootRew{\msym}}
\newcommand{\rtoe}{\rootRew{\esym}}
\newcommand{\rtoeabs}{\rootRew{\expoabs}} 
\newcommand{\rtoevar}{\rootRew{\expovar}} 
\newcommand{\esym}{{\mathsf e}}

\newcommand{\msym}{\mathsf{m}}

\newcommand{\wsym}{{\mathsf{o}}} %Giulio
\newcommand{\wmsym}{{\wsym\msym}} %Giulio
\newcommand{\omsym}{{\wmsym}} %Giulio
\newcommand{\shufeqext}{\shufeqext} %Giulio
 \newcommand{\toe}{\Rew{\esym}}

\newcommand{\toeabs}{\Rew{\expoabs}}
\newcommand{\toevar}{\Rew{\expovar}}

\newcommand{\tomo}{\Rew{\omsym}}

\newcommand{\toeo}{\Rew{\wsym{\esym}}}

\newcommand{\tm}{t}
\newcommand{\tmtwo}{u}
\newcommand{\tmthree}{r}
\newcommand{\tmfour}{q}
\newcommand{\tmfive}{p}

\newcommand{\tmp}{\tm'}
\newcommand{\tmtwop}{\tmtwo'}

\newcommand{\var}{x}

\newcommand{\vartwo}{y}

\newcommand{\varthree}{z}

\newcommand{\varfour}{w}
\newcommand{\varfourp}{w'}

\newcommand{\varfirst}{\var_1}
\newcommand{\varsec}{\var_2}
\newcommand{\varthird}{\var_3}

\newcommand{\val}{v}
\newcommand{\valtwo}{\val'}

\newcommand{\ctxholep}[1]{\langle #1\rangle}
\newcommand{\ctxhole}{\ctxholep{\cdot}}

\newcommand{\sctx}{L}
\newcommand{\sctxtwo}{\sctx'}

\newcommand{\sctxp}[1]{\sctx\ctxholep{#1}}
\newcommand{\sctxtwop}[1]{\sctxtwo\!\ctxholep{#1}}

\newcommand{\arbctxp}[1]{\arbctxp{#1}}
\newcommand{\arbctxtwop}[1]{\arbctxtwop{#1}}

\ifthenelse{\boolean{acmartstyle}%\or \boolean{IEEEstyle}
}{
  
  }{
  
  }

\newcommand{\bad}{\bullet}
\newcommand{\good}{\circ}

\newcommand{\deriv}{d}
\newcommand{\derivtwo}{e}
\newcommand{\derivthree}{f}

\newcommand{\sizehole}[2]{|#2|_{#1}}

\newcommand{\sizemo}[1]{\sizehole{\osym\msym}{#1}}
\newcommand{\sizeeo}[1]{\sizehole{\osym\esym}{#1}}
\newcommand{\sizegco}[1]{\sizehole{\osym\gcsym}{#1}}
\newcommand{\sizemp}[1]{\sizehole{\osym\mpsym}{#1}}
\newcommand{\sizeep}[1]{\sizehole{\osym\epsym}{#1}}
\newcommand{\sizegcp}[1]{\sizehole{\osym\gcpsym}{#1}}
\ifthenelse{\boolean{IEEEstyle}}{
\usepackage{amssymb}
\usepackage{amsthm}
    \newtheorem{theorem}{Theorem}[section]
    \newtheorem{lemma}[theorem]{Lemma}
    \newtheorem{corollary}[theorem]{Corollary}
    \newtheorem{proposition}[theorem]{Proposition}
    \newtheorem{definition}[theorem]{Definition}
}{}

\ifthenelse{\boolean{easychairstyle}}{
\usepackage{amssymb}
\usepackage{amsthm}
    \newtheorem{theorem}{Theorem}[section]
    \newtheorem{lemma}[theorem]{Lemma}
    \newtheorem{corollary}[theorem]{Corollary}
    \newtheorem{proposition}[theorem]{Proposition}
    \newtheorem{definition}[theorem]{Definition}
}{}

\newcommand{\VSC}{\textnormal{VSC}\xspace}

\newcommand{\tow}{\Rew{\wsym}} %Giulio
\newcommand{\towngc}{\Rew{\wsym\neg\gcsym}}
\newcommand{\osym}{{\mathsf o}}

\newcommand{\la}[1]{\lambda #1.}

\newcommand{\myproof}[1]{
\ifthenelse{\boolean{omitproofs}}{\begin{IEEEproof} Proof available but omitted for readability. \end{IEEEproof}}{#1}}

\newcommand{\node}{\mathsf{n}}

\newcommand{\withproofs}[1]{\ifthenelse{\boolean{withproofs}}{#1}{}}
\newcommand{\withoutproofs}[1]{\ifthenelse{\boolean{withproofs}}{}{#1}}

\crefname{proposition}{Prop.}{Props.}
\crefname{theorem}{Thm.}{Thms.}
\crefname{lemma}{Lemma}{Lemmas}
\crefname{corollary}{Cor.}{Cors.}
\crefname{section}{Sect.}{Sects.}
\Crefname{section}{Section}{Sections}

\newcommand{\doubt}[1]{}

\newcommand{\letexp}{\mathsf{let}}
\newcommand{\letin}[3]{{\sf let}\ #1=#2\ {\sf in}\ #3}

\newcommand{\Rule}{\mathsf{r}}

\newcounter{numberone}
\newcounter{numberoneroman}
\newcounter{numberonealph}

\newcommand{\cbv}{CbV\xspace}

\newcommand{\ocbv}{Open \cbv}

\newcommand{\mytr}[1]{\underline{#1}}
\newcommand{\auxtr}[1]{\overline{#1}}

\makeatletter
\newcommand\Copy[2]{% #1 is a key, #2 is the text
        \marginpar{\scriptsize \ \ \hyperlink{hl-appendix-#1}{Proof p.\,{\pageref*{appendix-#1}}}}
	\immediate\write\@auxout{\unexpanded{\global\long\@namedef{mytext@#1}{#2}
  }}%
	#2%
}

\newcommand\Paste[1]{%
        \hypertarget{hl-appendix-#1}{}\label{appendix-#1}
	\renewcommand{\inappendix}[1]{}
	\ifcsname mytext@#1\endcsname
	\@nameuse{mytext@#1}%
	\else
	``??''
	\fi
	\renewcommand{\inappendix}[1]{#1}
}
\makeatother

\newcommand{\inappendix}[1]{#1}

\newcommand{\sizep}[2]{|#1|_{#2}}

\newcommand{\subpred}{\mathsf{sub}}
\newcommand{\notsubpred}{\mathsf{nsub}}
\newcommand{\issub}[1]{\subpred(#1)}

\newcommand{\subctx}{\sctx}

\newcommand{\subctxfirst}{\sctx_1}
\newcommand{\subctxsec}{\sctx_2}
\newcommand{\subctxthird}{\sctx_3}

\newcommand{\subctxp}[1]{\subctx\ctxholep{#1}}

\newcommand{\subctxfirstp}[1]{\subctxfirst\ctxholep{#1}}

\newcommand{\subctxthirdp}[1]{\subctxthird\ctxholep{#1}}

\newcommand{\subctxsecfp}[1]{\subctxsec\ctxholefp{#1}}

\renewcommand{\good}{useful\xspace}
\newcommand{\Good}{Useful\xspace}
\renewcommand{\bad}{non-useful\xspace}
\newcommand{\Bad}{Non-useful\xspace}

\newcommand\Crumb\mytr
\newcommand\CrumbAux\auxtr

\newcommand{\evarsym}{\esym_{\mathsf{var}}}
\newcommand{\eabssym}{\esym_{\mathsf{abs}}}
\renewcommand{\rtoevar}{\rootRew\evarsym}
\renewcommand{\toevar}{\Rew{\osym\evarsym}}
\renewcommand{\rtoeabs}{\rootRew\eabssym}
\renewcommand{\toeabs}{\Rew{\osym\eabssym}}

\newcommand{\gcvarsym}{\gcsym_{\mathsf{var}}}
\newcommand{\gcabssym}{\gcsym_{\mathsf{abs}}}

\makeatletter
\newcommand{\exder}{%
  \def\exderW[##1]{\triangleright_{##1}\ }%
  \def\exderWO{\triangleright\ }%
  \@ifnextchar[\exderW\exderWO%
  }
\makeatother

\newcommand\mydots{\hbox to .6em{.\hss.}}

\newcommand{\emssym}{\esym_\mathsf{abs}}
\newcommand{\gcsym}{\gc}

\newcommand{\rtoems}{\rootRew{\emssym}}
\newcommand{\rtogc}{\rootRew{\gc}}

\newcommand{\rtogcabs}{\rootRew{\gcabssym}}

\newcommand{\rtogcvar}{\rootRew{\gcvarsym}}

\newcommand{\pa}{polarity assignment\xspace}

\newcommand{\npa}{negative \pa}
\newcommand{\ppa}{positive \pa}
\newcommand{\LJ}{{\sl LJ}\xspace}
\newcommand{\LJF}{{\sl LJF}\xspace}
\newcommand{\Lpos}{$\lambda_{\symfont{pos}}$\xspace}
\newcommand{\Lopos}{$\lambda_{\osym\symfont{pos}}$\xspace}
\newcommand{\Loxpos}{\Loxposnospace\xspace}
\newcommand{\Loxposnospace}{$\lambda_{\osym\xsym\symfont{pos}}$}

\newcommand{\plc}{positive $\lambda$-calculus\xspace}

\newcommand{\psubctx}{M}
\newcommand{\psubctxtwo}{\psubctx'}
\newcommand{\psubctxp}[1]{\psubctx\ctxholep{#1}}
\newcommand{\psubctxtwop}[1]{\psubctxtwo\ctxholep{#1}}

\newcommand{\topos}{\Rew{\possym}}
\renewcommand{\topos}{\Rew{\osym_+}}
\newcommand{\xsym}{\symfont{x}}
\newcommand{\toepos}{\Rew{\osym\xsym_+}}
\newcommand{\toepngc}{\Rew{\osym\xsym_+\neg\gcsym}}

\newcommand{\ppctx}{N}
\newcommand{\ppctxp}[1]{\ppctx\ctxholep{#1}}

\newcommand{\ppsubctx}{N}
\newcommand{\ppsubctxp}[1]{\ppsubctx\ctxholep{#1}}
\newcommand{\ppsubctxtwo}{\ppsubctx'}
\newcommand{\ppsubctxtwop}[1]{\ppsubctxtwo\ctxholep{#1}}

\newcommand{\Lppos}{$\lambda_{\mathsf{pos'}}$\xspace}
\renewcommand{\Lppos}{\Loxpos}

\newcommand{\posmsym}{\mpsym}
\newcommand{\posesym}{\epsym}

\newcommand{\toposm}{\Rew{\posmsym}}
\renewcommand{\toposm}{\tomp}

\newcommand{\topose}{\Rew{\posesym}}

\renewcommand{\topose}{\toep}

\newcommand{\ppossym}{\mathsf{pos'}}
\renewcommand{\ppossym}{\epossym}

\newcommand{\topposn}{\Rewn{\ppossym}}

\newcommand{\sizeposm}[1]{\sizehole{\ompsym}{#1}}

\newcommand{\epossym}{\osym\xsym_+}

\newcommand{\isubs}{\sigma}

\newcommand{\isubempty}{\cdot}

\newcommand{\vtopptm}[1]{{\color{blue}[}#1{\color{blue}]}}
\newcommand{\vtoppctx}[1]{\vtopptm{#1}}
\renewcommand{\vtopptm}[1]{\semb{#1}}

\newcommand{\egoodsym}{\esym_\mathsf{good}}
\newcommand{\egoodasym}{\esym_\mathsf{good_1}}
\newcommand{\egoodbsym}{\esym_\mathsf{good_2}}
\newcommand{\ebadsym}{\esym_\mathsf{bad}}

\newcommand{\rtoegooda}{\rootRew{\egoodasym}}
\newcommand{\rtoegoodb}{\rootRew{\egoodbsym}}
\newcommand{\toegood}{\Rew{\osym\egoodsym}}
\newcommand{\toegooda}{\Rew{\osym\egoodasym}}
\newcommand{\toegoodb}{\Rew{\osym\egoodbsym}}

\newcommand{\toebad}{\Rew{\osym\ebadsym}}

\newcommand{\toevaro}{\Rew{\osym\evarsym}}
\renewcommand{\toevar}{\toevaro}
\newcommand{\togco}{\Rew{\osym\gcsym}}

\makeatletter
\newsavebox{\@brx}
\newcommand{\llangle}[1][]{\savebox{\@brx}{\(\m@th{#1\langle}\)}%
  \mathopen{\copy\@brx\kern-0.5\wd\@brx\usebox{\@brx}}}
\newcommand{\rrangle}[1][]{\savebox{\@brx}{\(\m@th{#1\rangle}\)}%
  \mathclose{\copy\@brx\kern-0.5\wd\@brx\usebox{\@brx}}}
\makeatother

\newcommand{\goodpred}{\mathsf{good}}
\newcommand{\isgood}[1]{\goodpred(#1)}
\newcommand{\goodctx}{G}

\newcommand{\goodctxfp}[1]{\goodctx\ctxholefp{#1}}

\newcommand{\badpred}{\mathsf{bad}}
\newcommand{\isbad}[1]{\badpred(#1)}
\newcommand{\badctx}{B}

\newcommand{\togoodvscn}{\Rewn{\goodvscsym}}

\newcommand{\octx}{O}
\newcommand{\octxp}[1]{\octx\ctxholep{#1}}
\newcommand{\octxtwo}{\octx'}

\newcommand{\octxfirst}{\octx_1}
\newcommand{\octxfirstp}[1]{\octxfirst\ctxholep{#1}}
\newcommand{\octxsec}{\octx_2}

\newcommand{\ctxholefp}[1]{\llangle #1\rrangle}

\newcommand{\octxfp}[1]{\octx\ctxholefp{#1}}

\newcommand{\asym}{a}

\newcommand{\symfont}[1]{\mathtt{#1}}
\newcommand{\vscsym}{\symfont{vsc}}
\newcommand{\mpsym}{\msym_+}
\newcommand{\epsym}{\esym_+}
\newcommand{\mepsym}{\esym\msym\esym_+}
\newcommand{\gcpsym}{\gcsym_+}
\newcommand{\ompsym}{\osym\mpsym}
\newcommand{\oepsym}{\osym\epsym}
\newcommand{\omepsym}{\osym\mepsym}
\newcommand{\ogcpsym}{\osym\gcpsym}

\newcommand{\toovsc}{\Rew{\osym\vscsym}}
\newcommand{\rtomep}{\rootRew{\mepsym}}
\newcommand{\tomep}{\Rew{\omepsym}}
\newcommand{\rtomp}{\rootRew{\mpsym}}
\newcommand{\tomp}{\Rew{\ompsym}}
\newcommand{\rtoep}{\rootRew{\epsym}}
\newcommand{\toep}{\Rew{\oepsym}}
\newcommand{\rtogcp}{\rootRew{\gcpsym}}
\newcommand{\togcp}{\Rew{\ogcpsym}}

\newcommand{\lvsc}{$\l_{\osym\vscsym}$\xspace}

\newcommand{\ans}{a}
\newcommand{\anstwo}{\ans'}

\newcommand{\myparagraph}[1]{\bigskip 

\textit{#1.}}

\newcommand{\myparagraphmed}[1]{\medskip 

\textit{#1.}}

\newcommand{\myparagraphsmall}[1]{\smallskip 

\textit{#1.}}

\newcommand{\coresym}{\symfont{core}}
\newcommand{\tocore}{\Rew{\osym\coresym}}

\newcommand{\toone}{\Rew{1}}
\newcommand{\totwo}{\Rew{2}}

\newcommand{\aanswer}{almost answer\xspace}
\newcommand{\aofv}[1]{\symfont{aofv}(#1)}
\newcommand{\ofv}[1]{\symfont{ofv}(#1)}
\newcommand{\ntm}{n}
\newcommand{\ntmtwo}{\ntm'}

\newcommand{\forms}{\Gamma}
\newcommand{\typtm}{A}
\newcommand{\typtmtwo}{B}
\newcommand{\typtmthree}{C}

\setboolean{withproofs}{false}

\newboolean{appendix}
\setboolean{appendix}{false}
\usepackage{appendix}
\usepackage{cleveref}
\ifthenelse{\boolean{mfpsstyle}}{\capturecounter{thm}}{\capturecounter{theorem}}
\usepackage{marginnote}

\newcommand{\NoteProof}[1]{
	\ifthenelse{\boolean{appendix}}{
	\marginnote{Link to the original position of \cref{#1}, p. \pageref{#1}}
	}{
	\marginnote{{Proof\,p.\,{\pageref{app:#1}}}}
	}
}

\usepackage{bussproofs}
\renewcommand{\goodctx}{U}
\renewcommand{\ppsubctx}{E}
\renewcommand{\badctx}{N}
\renewcommand{\ppctx}{E}
\renewcommand{\psubctx}{E}

\renewcommand{\NoteProof}[1]{\withproofs{\ifthenelse{\boolean{appendix}}{, originally at p. \pageref{#1}}{, proof at p.\,{\pageref{app:#1}}}}}

\newcommand{\NoteProofNoComma}[1]{\withproofs{\ifthenelse{\boolean{appendix}}{Originally at p. \pageref{#1}}{Proof at p.\,{\pageref{app:#1}}}}}

\renewcommand{\goodpred}{\mathsf{usef}}
\renewcommand{\badpred}{\mathsf{nusef}}

\renewcommand{\egoodsym}{\esym_\mathsf{u}}
\renewcommand{\egoodasym}{\esym_\mathsf{u_1}}
\renewcommand{\egoodbsym}{\esym_\mathsf{u_2}}
\renewcommand{\ebadsym}{\esym_\mathsf{nu}}

\renewcommand{\togoodvscn}{\tocore^*}

\renewcommand{\LJF}{LJF\xspace}

\usepackage{enumitem}
\setlistdepth{5}

\usepackage{enticsmacro}

\def\conf{MFPS 2024}
\def\myconf{mfps40} 	%%Fill in the acronym for your conference (with year)
\volume{4}			%Fill in the ENTICS volume number here
\def\volu{4}			% and here
\def\papno{3}			%Fill in your paper number here
\def\mydoi{14758}%
\def\lastname{Accattoli and Wu} %Lastnames appear in the running header 
\def\runtitle{Positive Focusing is Directly Useful} %Short title appears in the running header on even pages. 
\def\copynames{B. Accattoli, J.-H. Wu}    %% Fill in the first initial and last name of the authors
\begin{document}
%%%Note the beginning and end of the frontmatter section that starts here%%%%%
\begin{frontmatter}
  \title{Positive Focusing is Directly Useful} 						%%Title here and the
% \thanks[ALL]{General thanks to everyone who should be thanked.}   %%Text of \thanks[ALL} here..
 %%%%%%%%%%%%%%%%%%%%%%%%%%%%			This Thanks is optional.
  %%%%Now the author(s) names(s)%%%%%
  \author{Beniamino Accattoli\thanksref{myemail}}	%%Note NO SPACE between 
   \author{Jui-Hsuan Wu\thanksref{coemail}}		%last name and \thanksref{...} 
    %%%Next come the addresses%%%%
   \address{Inria \& LIX, Ecole Polytechnique, UMR 7161, Palaiseau, France}  							
   \thanks[myemail]{Email: \href{mailto:beniamino.accattoli@inria.fr} {\texttt{\normalshape beniamino.accattoli@inria.fr}}} 
   %%%Note: if both authors share same institution, only list the address once, after the second 
   %%%author. 
   %%%There also is a link from the first author to the co-author's address to show how to list 
   %%%affiliations to more than one institution, when needed. 
  %\address[b]{My Co-author's Department\\My Co-author's University\\
   % My Co-author's City, My Co-author's Country} 
  \thanks[coemail]{Email:  \href{mailto:jwu@lix.polytechnique.fr} {\texttt{\normalshape       jwu@lix.polytechnique.fr}}}
\begin{abstract} 
Recently, Miller and Wu introduced the \emph{positive $\lambda$-calculus}, a call-by-value $\lambda$-calculus with sharing obtained by assigning proof terms to the positively polarized focused proofs for minimal intuitionistic logic. The positive $\lambda$-calculus stands out among $\l$-calculi with sharing for a \emph{compactness} property related to the sharing of variables. We show that\textemdash thanks to compactness\textemdash the positive calculus neatly captures the core of \emph{useful sharing}, a technique for the study of reasonable time cost models.
\end{abstract}
\end{frontmatter}

%\section{Introduction}\label{intro}
%%%%%%%%%%%%%%%%%%%%%%
%\input{01-Introduction}
%%%%%%%%%%%%%%%%%%%%%%
\section{Introduction}
\label{sect:intro}
Andreoli's focusing is a technique for restricting the space of proofs to a subset with good structural properties, called \emph{focused proofs} \cite{DBLP:journals/logcom/Andreoli92}. Notably, it helps in the \emph{proof search} approach to computation. 

Focused proof systems have quickly become a widespread tool in proof theory and---via the Curry-Howard isomorphism---in the study of $\l$-calculi. They are used well beyond proof search, for instance, in connection to pattern matching by Zeilberger \cite{DBLP:conf/popl/Zeilberger08} and Krishnaswami \cite{DBLP:conf/popl/Krishnaswami09}, proof nets and expansion proofs by Chaudhuri et al. \cite{DBLP:conf/ifipTCS/ChaudhuriMS08,DBLP:journals/logcom/ChaudhuriHM16}, synthetic connectives by Chaudhuri \cite{DBLP:conf/lpar/Chaudhuri08},  abstract machines by Brock-Nannestad et al. \cite{DBLP:conf/ppdp/Brock-Nannestad15}, decidability of contextual equivalence for sum types by Scherer and R\'emy \cite{DBLP:conf/icfp/SchererR15} and Scherer \cite{DBLP:conf/popl/Scherer17}, contextual equivalence by Rioux and Zdancewic \cite{DBLP:journals/pacmpl/RiouxZ20}, synthetic inference rules by Marin et al. \cite{DBLP:journals/apal/MarinMPV22}, refinement types by Economou et al. \cite{DBLP:journals/toplas/EconomouKD23}, and certainly in even more studies. The present paper adds one more instance, studying a connection between focusing and sharing for $\l$-terms.

\myparagraph{Positive Focusing and $\l$-Terms} 
In \cite{DBLP:conf/csl/0001W23}, Miller and Wu show how the focused intuitionistic
proof system \LJF  by Liang and Miller \cite{DBLP:journals/tcs/LiangM09} can be used to design term structures. 
 In \LJF, formulas are polarized. A key theorem of \LJF
is that different polarizations do not affect provability: if a formula is
provable (resp. not provable) in \LJ, then the formula is provable (resp. not
provable) in \LJF with \emph{any} polarization. Different polarizations of a
provable formula, however, induce focused proofs of very different shapes. %By annotating proofs with proof terms, it is then possible to obtain different calculi. 

Via annotations of proofs with proof terms, Miller and Wu study the shape of polarized proofs of \emph{minimal} intuitionistic logic, that is, for the logic of implications, which is the basic setting of the Curry-Howard isomorphism. %Since implications can only be polarized negatively, a polarization is equivalent to a \pa to atomic formulas. 
They consider the two natural uniform polarizations having either all atomic formulas polarized negatively, or positively (non-atomic formulas are all negative, because implication is negative). Two very different term structures arise. The \npa yields the usual $\l$-calculus. The \ppa, instead, yields a quite different syntax, with a restricted form of application and accounting for sharing via a notion of explicit substitution.

\myparagraph{The Positive $\l$-Calculus} Based on this positively-polarized syntax, Wu introduced the \emph{\plc}  \Lpos \cite{DBLP:conf/aplas/Wu23}, a calculus with explicit substitutions endowed with call-by-value (shortened to \cbv) evaluation, because substituting applications (as in call-by-name) would break the shape of positive terms. 

The \plc is an unusual calculus and the aim of this paper is to show its relevance. At first sight, it is yet another sharing-based presentations of \cbv evaluation. It stands out, however, for a peculiar treatment of variables in relation to sharing, here dubbed \emph{compactness}, that to our knowledge is new. The main contribution of this paper is to show that compactness considerably simplifies the definition and the study of \emph{useful sharing}, a form of shared evaluation first introduced by Accattoli and Dal Lago \cite{DBLP:journals/corr/AccattoliL16} to study reasonable time cost models for the $\l$-calculus.

In order to properly explain the novelty of the positive $\l$-calculus, we are now going to outline three concepts: sharing-based presentations of \cbv evaluation, the sharing of variables, and useful sharing.

\myparagraph{700 Sharing-Based Presentations of \cbv Evaluation} In the literature, sharing is an overloaded word. The most basic form of sharing is \emph{sub-term sharing}, which can be obtained by simply adding to the standard syntax of the $\l$-calculus a $\letin \var\tmtwo\tm$ construct, which shares $\tmtwo$ between all the occurrences of $\var$ in $\tm$. The use of $\letexp$-expressions is pervasive in presentations of \cbv $\l$-calculi, typically in Moggi \cite{Moggi88tech,DBLP:conf/lics/Moggi89}. In $\letin \var\tmtwo\tm$, it is usually assumed that $\tmtwo$ shall be evaluated before $\tm$, but such an assumption is often dropped when studying sharing. Additionally, $\letin \var\tmtwo\tm$ is often rather more concisely noted as an explicit substitution $\tm\esub\var\tmtwo$ (in this paper, meta-level substitution is noted $\tm\isub\var\tmtwo$).

In a \cbv setting, having both explicit substitutions $\tm\esub\var\tmtwo$ and applications $\tm\tmtwo$ is somewhat redundant, as explicit substitutions can be used to constrain the shape of applications. It is possible, indeed, to have only values $\val$ for one or both sub-terms of applications $\tm\tmtwo$, that is, they can be constrained to be of shape $\val\tmtwo$ (or $\tm\val$, or even $\val\valtwo$). The idea is to apply a transformation $\vtopptm\cdot$ turning $\tm\tmtwo$ into $(\var\vtopptm\tmtwo)\esub\var{\vtopptm\tm}$ with $\var$ fresh (or into $(\vtopptm\tm\var)\esub\var{\vtopptm\tmtwo}$, or  $(\var\vartwo)\esub\var{\vtopptm\tm}\esub\vartwo{\vtopptm\tmtwo}$ with $\vartwo$ fresh). It is typical of \cbv (rather than call-by-name), because, for the restriction to be stable by reduction, it should be forbidden to substitute applications. Instances of \cbv calculi with restrained applications abound, two notable examples being the calculus of \emph{A-normal forms} by Sabry and Felleisen \cite{DBLP:conf/lfp/SabryF92,DBLP:conf/pldi/FlanaganSDF93} and the \emph{fine-grained} \cbv calculus by Levy et al. \cite{DBLP:journals/iandc/LevyPT03}. Applications are also restricted in Sestoft's study of call-by-need \cite{DBLP:journals/jfp/Sestoft97}, or that of Walker's on substructural type systems \cite{WalkerSubstructural}. 

We shall follow Accattoli et al. \cite{DBLP:conf/ppdp/AccattoliCGC19} and refer to these decompositions of applications as to \emph{crumbled} calculi. By tweaking the rewriting rules, one can also further restrict the immediate sub-terms of applications to be variables (rather than values). This gives rise to at least nine different \cbv calculi, one with ordinary application $\tm\tmtwo$ and eight crumbled variants $\val\tmtwo$, $\var\tmtwo$, $\tm\valtwo$, $\val\valtwo$, $\var\valtwo$, $\tm\vartwo$, $\val\vartwo$, and $\var\vartwo$. One can also decide at least (each choice doubling the number of possible presentations): 
\begin{itemize}
\item \emph{Nested $vs$ flattened ES}: whether explicit substitutions can be nested (as in $\tm\esub\var{\tmtwo\esub\vartwo\tmthree}$) or have to be flattened (as in $\tm\esub\var{\tmtwo}\esub\vartwo\tmthree$); 
\item \emph{Granularity of substitutions}: whether evaluation is small-step (substitution acts on all occurrences of a variable at once) or micro-step (variable occurrences are replaced one at a time); 
\item \emph{Variables $vs$ values}: whether variables are values, and thus can trigger substitution redexes, or not, that is, only abstractions can be substituted.
\end{itemize}
None of these choices affects the expressivity of the calculus, but the various calculi have different properties and degrees of manageability. In particular, the proof that a crumbled calculus is as expressive as the full one depends on the choices, and the proof might be non-trivial, as we shall see.

According to this classification, Wu's positive $\l$-calculus is a crumbled calculus with applications of shape $\var\vartwo$ (the minimalistic one), flattened substitutions, micro-step, and where variables are not values. It distinguishes itself from all these calculi, however, because it also has a new \emph{compactness} feature that %somewhat brings the removal of variables from values a step further, and 
solves a pervasive issue of sub-term sharing.

\myparagraph{Sharing of Variables and Compactness} In $\l$-calculi with sharing, usually variables can be shared, that is $\tm\esub\var\vartwo$ is a valid term. This leads to the possibility of having \emph{renaming chains}, such as:
\begin{equation}
\tm\esub{\var_1}{\var_2}\esub{\var_2}{\var_3}\ldots \esub{\var_{n-1}}{\var_n}
\label{eq:ren-chain}
\end{equation}
These chains are an issue because they lead to both space and time inefficiencies. Optimizations to prevent their creation are adopted for instance by Sands et al. \cite{DBLP:conf/birthday/SandsGM02}, Wand \cite{DBLP:journals/lisp/Wand07}, Friedman et al. \cite{DBLP:journals/lisp/FriedmanGSW07}, and Sestoft \cite{DBLP:journals/jfp/Sestoft97}. The first systematic study of renaming chains is by Accattoli and Sacerdoti Coen \cite{DBLP:journals/iandc/AccattoliC17}, with respect to time, where they show that it is enough to remove variables from values to avoid time inefficiencies related to renaming chains. Recently, Accattoli et al. have shown that the dynamic removal of renaming chains is essential for the only known reasonable notion of logarithmic space in the $\l$-calculus \cite{DBLP:conf/lics/AccattoliLV22}.%In particular, they show that a way to avoid the creations of such chains is to define values as being only abstractions (as it is done in most implementative studies), thus forbidding the substitution of variables, instead of values being both variables and abstractions (as defined originally by Plotkin \cite{DBLP:journals/tcs/Plotkin75} and as done by most theoretical studies).

The new feature of the positive $\l$-calculus \Lpos is that \emph{it does not share variables}, that is, $\tm\esub\var\vartwo$ does \emph{not} belong to the syntax of \Lpos. This fact removes the issue of renaming chains altogether, with no need to design optimizations to prevent their creations, or removing variables from values, because renaming chains simply \emph{cannot be expressed}.

In fact, \Lpos pushes things one step further, forging a sort of syntactical duality with respect to sharing: 
\begin{itemize}
\item Variables cannot appear in explicit substitutions and, additionally, are the only constructors beside explicit substitutions;
\item Dually, abstractions and applications appear in explicit substitutions but \emph{not} out of them.
\end{itemize}
That is, $\var$ is a positive term but, for instance, $\var\vartwo$ and $\la\var\var$ are not positive terms, only $\varthree\esub\varthree{\var\vartwo}$ and $\varthree\esub\varthree{\la\var\var}$ (or more generally $\tmtwo\esub\varthree{\var\vartwo}$ and $\tmtwo\esub\varthree{\la\var\var}$, where $\tmtwo$ is a positive term) are positive terms. This removes the further issue of whether $\var\vartwo$ and $\varthree\esub\varthree{\var\vartwo}$ are distinct terms, which is relevant in the study of \cbv program equivalences, see Accattoli et al. \cite{DBLP:journals/corr/abs-2303-08161}. One might argue that a similar approach to applications is already present in calculi related to the sequent calculus (in the style of Curien and Herbelin \cite{DBLP:conf/icfp/CurienH00}), but to our knowledge the extension of this approach to abstractions is new, and relevant for useful sharing, see \refsect{dissecting}.

We dub \emph{compactness} the fact that variables are not shared and are the only terms out of ESs. With compact sub-term sharing, terms have shape $\var_1\esub{\var_1}{\tmtwo_1}\ldots\esub{\var_n}{\tmtwo_1}$, where each of the $\tmtwo_1$,\ldots, $\tmtwo_n$ is an abstraction or an application (which do not belong to the grammar of terms).

\myparagraph{Useful Sharing} Our aim here is promoting the compactness of \Lpos by showing its impact on \emph{useful sharing}, a concept apparently unrelated to focusing. Useful sharing is a technique introduced by Accattoli and Dal Lago \cite{DBLP:journals/corr/AccattoliL16}, and then refined in call-by-value by Accattoli et al. \cite{DBLP:conf/lics/AccattoliCC21}, to study reasonable time cost models for $\l$-calculi. It works at the level of micro-step evaluation, that is, of replacements of single variable occurrences. The basic idea is that (in \cbv) one should replace a variable occurrence with a copy of a shared abstraction only when it is useful to create $\beta$-redexes, that is, in the following case:
\begin{equation}
(\var\tm)\esub\var{\la\vartwo\tmtwo} \ \ \to\ \  ((\la\vartwo\tmtwo)\tm)\esub\var{\la\vartwo\tmtwo}
\label{eq:imm-useful}
\end{equation}
While one should avoid replacements that do not create $\beta$-redexes, i.e. are not useful, as the following one:
\begin{equation}
(\tm\var)\esub\var{\la\vartwo\tmtwo} \ \ \to\ \  (\tm (\la\vartwo\tmtwo))\esub\var{\la\vartwo\tmtwo}
\label{eq:non-useful}
\end{equation}
Avoiding non-useful replacements leads to considerable speed-ups, that can even be \emph{exponential} for some terms, in the case of strong evaluation (that is, when evaluation goes under abstraction) \cite{DBLP:journals/corr/AccattoliL16,DBLP:conf/lics/AccattoliCC21}.

While the intuition behind useful sharing is easy to convey, the technical details are complex, because cases \refeq{imm-useful} and \refeq{non-useful} are not the only possible ones, and various complications arise. A first simplified setting for useful sharing is given by crumbled $\l$-calculi where arguments can only be variables (that is, with applications of shape $\tm\vartwo$, $\val\vartwo$, or $\var\vartwo$), since then non-useful replacements such as those in \refeq{non-useful} cannot be expressed and are then  ruled out. This is already known and used by Accattoli et al. \cite{DBLP:conf/lics/AccattoliCC21}. 

One of the complex aspects of useful sharing is related to renaming chains (as in \refeq{ren-chain}). They force the distinction between steps as in \refeq{imm-useful}, which are \emph{directly useful}, and steps over a renaming chain such as:
\begin{equation}
(\var_1\tm)\esub{\var_1}{\var_2}\ldots \esub{\var_{n-1}}{\var_n}\esub{\var_n}{\la\vartwo\tmtwo} \ \ \to\ \  (\var_1\tm)\esub{\var_1}{\var_2}\ldots \esub{\var_{n-1}}{\la\vartwo\tmtwo}\esub{\var_n}{\la\vartwo\tmtwo}
\label{eq:rel-useful}
\end{equation}
These replacements are \emph{indirectly useful}: they do not directly create a $\beta$-redex and yet they contribute to the \emph{future} creation of $\beta$-redexes. Continuing with evaluation, indeed, $\la\vartwo\tmtwo$ shall replace the content of all the explicit substitutions in the chain and finally be substituted for $\var_1$, creating a $\beta$-redex. Indirectly useful steps cannot be avoided, otherwise some $\beta$-redexes are never created, and evaluation gets stuck.

Thanks to compactness, the positive $\l$-calculus \Lpos has no renaming chains and thus indirectly useful replacements are simply \emph{ruled out}. Evaluation is not stuck in \Lpos, though, because indirectly useful steps somehow transform into directly useful steps in \Lpos, as explained in \refsect{core-factorization}. Since \Lpos also has minimalistic applications of shape $\var\vartwo$, non-useful replacements are ruled out as well. Therefore, \Lpos has only useful replacements, and only the \emph{directly} useful ones (hence the title). Essentially, \Lpos captures the \emph{core} of useful sharing, ruling out the technicalities. %What is surprising is that the inception of \Lpos is in the study of focusing, and not of useful sharing.

\myparagraph{Open \cbv} Concretely, we develop our study with respect to Accattoli and Guerrieri's framework of \emph{Open \cbv} \cite{DBLP:conf/aplas/AccattoliG16}, that is, we consider \emph{weak evaluation} (i.e. not under abstraction) and \emph{possibly open terms}. It is an intermediate setting between:
\begin{enumerate}
\item Weak evaluation and closed terms, where many subtleties of useful sharing are not observable, and 
\item Strong evaluation (which implies dealing with open terms), where further technicalities arise. 
\end{enumerate}
The same approach is followed by other works on useful sharing by Accattoli and co-authors \cite{DBLP:conf/lics/AccattoliC15,DBLP:journals/scp/AccattoliG19,DBLP:conf/csl/AccattoliL22}---see in particular the introduction of \cite{DBLP:conf/csl/AccattoliL22} for an extensive discussion of this choice.

\myparagraph{Architecture of Our Study} After all these preliminary explanations, we can now explain our results. As a reference for a sharing-based $\l$-calculus for \cbv, we take (a micro-step presentation of) Accattoli and Paolini's \emph{value substitution calculus} (VSC) \cite{DBLP:conf/flops/AccattoliP12}, which is also the reference for the study of \cbv reasonable time and useful sharing by Accattoli et al. \cite{DBLP:conf/lics/AccattoliCC21}. The VSC has (possibly nested) explicit substitutions and ordinary applications of shape $\tm\tmtwo$. We then define (our slight variant of) Wu's positive $\l$-calculus \Lpos. 

Intuitively, the paper is devoted to proving the equivalence of the Open \Lpos and the Open VSC. Often, the equivalence of two calculi is stated as a bisimulation property. For crumbled calculi ruling out some steps\textemdash as it is the case for the positive $\l$-calculus\textemdash this is not possible, because the ruled out steps cannot be simulated. A finer approach is needed.

We thus define a \emph{core sub-calculus} of the Open VSC, which has only directly useful replacements, and prove a \emph{factorization theorem} for the Open VSC, stating that every reduction sequence can be factored into a core one followed by a non-core one. Moreover, we provide a theorem ensuring that the Open VSC and the Core Open VSC are termination equivalent. 

Next, we define a translation $\vtopptm\cdot$ from the VSC to \Lpos and show that:
\begin{enumerate}
\item \emph{Simulation}: it induces a simulation of the Core Open VSC by the Open \Lpos, and
\item \emph{Normal forms}: it sends core normal forms into appropriate normal forms of \Lpos.
\end{enumerate}
From these properties, it follows that the Core Open VSC  and the Open \Lpos are termination equivalent. 

This proof technique is an original contribution of the paper. In particular, it is the first time that usefulness is justified via a factorization theorem.

Lastly, we show that the compactness of \Lpos induces a further relevant property: its open evaluation has the \emph{diamond property}, a strong form of confluence, which is not true for the (micro-step) Open VSC.

\myparagraph{Proofs} Proofs are in the technical report \cite{accattoli:hal-04606194}. %In case of acceptance, this version with Appendix will be uploaded on ArXiv.

%%%%%%%%%%%%%%%%%%%%%%
%\input{02-Rewriting_Preliminaries}
%%%%%%%%%%%%%%%%%%%%%%
\section{Preliminaries: Rewriting Notions and Notations}
\label{sect:rewriting-notions}
Given a rewriting relation $\Rew{\Rule}$, we write $\deriv \colon \tm \Rew{\Rule}^* \tmtwo$ for a $\Rew{\Rule}$-reduction sequence from $\tm$ to $\tmtwo$, the length of which is noted $\size{\deriv}$. Moreover, we use $\size{\deriv}_a$ for the number of $a$-\emph{steps} in $\deriv$, for a sub-relation $\Rew{a}$ of $\Rew{\Rule}$. 

%A term $\tm$ is $\Rule$-\emph{normal} if there is no $\tmtwo$ such that $\tm \Rew{\Rule} \tmtwo$.
%An evaluation $\deriv \colon \tm \Rew{\Rule}^* \tmtwo$ is \emph{$\Rule$-normalizing} if $\tmtwo$ is $\Rule$-normal.
A term $\tm$ is \emph{weakly $\Rule$-normalizing} if $\deriv \colon \tm \Rew{\Rule}^* \tmtwo$ with $\tmtwo$ $\Rule$-normal; and $\tm$ is \emph{strongly $\Rule$-normalizing} if there are no diverging $\Rule$-sequences from $\tm$, or, equivalently, if all its $\Rule$-reducts are strongly $\Rule$-normalizing.

Given two reductions $\toone$ and $\totwo$ we use $\toone\totwo$ for their composition, defined as $\tm \toone \totwo \tmtwo$ if $\tm \toone\tmthree\totwo \tmtwo$ for some $\tmthree$. We also write, \eg, $\toone\totwo \subseteq \Rewn{3}$ to state that $\tm \toone \totwo \tmtwo$ implies $\tm \Rewn{3} \tmtwo$

\myparagraph{Diamond Property} According to Dal Lago and Martini \cite{DBLP:journals/tcs/LagoM08}, a relation $\Rew{\Rule}$ is \emph{diamond} if $\tmtwo_1 \,{}_\Rule\!\!\lto \tm \Rew{\Rule} \tmtwo_2$ and $\tmtwo_1 \neq \tmtwo_2$ imply $\tmtwo_1 \Rew{\Rule} \tmthree \, {}_\Rule\!\!\lto \tmtwo_2$ for some $\tmthree$. 
	%Terminology in the literature is inconsistent: Terese \cite[Exercise 1.3.18]{Terese} dubs this property $\textup{CR}^1$, and defines the diamond more restrictively, without requiring $u_1 \neq u_2$ in the hypothesis: $u_1$ and $u_2$ have to join even if $u_1 = u_2$.
%\footnotetext{\giulio{Terese \cite{Terese} defines the diamond without requiring $u_1 \neq u_2$, and dubs our diamond~$\textup{CR}^1$.}}
%Dal Lago and Martini show that 
If $\Rew{\Rule}$ is diamond then:
\begin{enumerate}
	\item \emph{Confluence}:
	$\Rew{\Rule}$ is confluent, that is, $\tmtwo_1 \,{}_\Rule^*\!\!\lto \tm \Rew{\Rule}^* \tmtwo_2$  implies $\tmtwo_1 \Rew{\Rule}^* \tmthree \, {}_\Rule^*\!\!\lto \tmtwo_2$ for some $\tmthree$; 
%	\item any term $\tm$ has at most \emph{one normal form} (\ie if $\tm \Rew{\Rule}^* \tmtwo$  and $\tm \Rew{\Rule}^* \tmthree$ with $\tmtwo$ and $\tmthree$ $\Rule$-normal, then $\tmtwo = \tmthree$); \ben{Questo secondo punto mi sembra inutile, tutti sanno che la confluenza impliza l'unicit\'a della forma normale.}
	\item \emph{Length invariance}: all $\Rule$-reduction sequences to normal form with the same start term have the same length (\ie if $\deriv \colon \tm \Rew{\Rule}^* \tmtwo$  and $\deriv' \colon \tm \Rew{\Rule}^* \tmtwo$ with $\tmtwo$ $\Rew{\Rule}$-normal then $\size{\deriv} = \size{\deriv'}$);
	\item \emph{Uniform normalization}: $\tm$ is weakly $\Rule$-normalizing if and only if it is strongly $\Rule$-normalizing.
\end{enumerate}
Basically, the diamond property captures a more liberal form of determinism.

%%%%%%%%%%%%%%%%%%%%%%
%\input{03-micro-vsc}
%%%%%%%%%%%%%%%%%%%%%%
\section{The (Micro-Step) Open Value Substitution Calculus}
\label{sect:vsc}
%%%%%
\begin{figure}[t!]
\centering
\arraycolsep=3pt
\tabcolsep=3pt
\fbox{
\begin{tabular}{c}
				$\begin{array}{rrll| rrll}
				\multicolumn{4}{c}{\textsc{Language}}
				&
				\multicolumn{4}{c}	{\textsc{Contexts}}
				\\[4pt]
					\textsc{Terms} &  \tm,\tmtwo,\tmthree,\tmfour,\tmfive & \grameq& \val \mid \tm\tmtwo \mid \tm\esub\var\tmtwo\phantom{\hspace{.3cm}}
					&
					\quad\textsc{Subst.} & \sctx,\sctxtwo & \grameq & \ctxhole \mid \sctx\esub\var\tmtwo
					\\
					\textsc{Values} & \val,\valtwo & \grameq & \var \mid \la\var\tm\phantom{\hspace{.3cm}}
					&
					\quad\textsc{Open} & \octx,\octxtwo & \grameq & \ctxhole \mid \octx\tm \mid \tm\octx \mid \tm\esub\var\octx \mid \octx\esub\var\tmtwo
					\\
					\textsc{Answers} & \ans,\anstwo & \grameq & \sctxp{\la\var\tm}
				\end{array}$
			\\[37pt]
			\hline

			$\begin{array}{c@{\hspace{0.25cm}}|@{\hspace{0.25cm}}c}
			\multicolumn{2}{c}{\textsc{Reduction rules}}
			\\[4pt]
			\begin{array}{rrll}
			\textsc{Multiplicative} &\sctxp{\la\var\tm}\tmtwo & \rtom & \sctxp{\tm\esub\var\tmtwo}
			\\
			\textsc{Exponential} &\octxfp{\var}\esub\var{\subctxp{\val}} &
			\rtoe & \subctxp{\octxfp{\val}\esub{\var}{\val}} 
			\\
			\textsc{Garbage coll.} &\tm\esub{\var}{\subctxp{\val}} &
			\rtogc & \subctxp\tm \quad \text{ if } \var \notin \fv\tm
			\\[4pt]
			\hline
			\textsc{Notation} & \toovsc &   \defeq & \Rew{\wsym\msym} \cup \Rew{\wsym\esym}\cup \Rew{\wsym\gcsym}
			\end{array}
					& 
					\begin{array}{c}
					\textsc{Contextual closure}
					\\[4pt]					
					\AxiomC{$\tm \rootRew{\asym} \tm'$}
					\AxiomC{$\asym{\in}\{\msym,\esym,\gcsym\}$}
					\BinaryInfC{$\octxp\tm \Rew{\wsym\asym} \octxp{\tm'}$}
					\DisplayProof
					\end{array}
			\\[5pt]

			\end{array}$
				\end{tabular}
				}
				\caption{The Open (Micro-Step) Value Substitution Calculus \lvsc.}
				\label{fig:weak-vsc-def}	
		\end{figure}
%%%
In this section, we present our system of reference for \ocbv, Accattoli and Paolini's \emph{value substitution calculus} \cite{DBLP:conf/flops/AccattoliP12} (shortened to VSC). We shall adopt a micro-step presentation (explained below) of the open fragment, that is, of weak evaluation on possibly open terms. This variant is defined in \reffig{weak-vsc-def} and shall be denoted with \lvsc. We also recall some of its basic properties, which shall either be used in the next section or used to compare \lvsc with the positive $\l$-calculus.

\myparagraph{Terms} The \VSC is a \cbv $\lambda$-calculus extended with $\letexp$-expressions, similarly to Moggi's \cbv calculus \cite{Moggi88tech,DBLP:conf/lics/Moggi89}. 
We do however write a $\letexp$-expression $\letin\var\tmtwo\tm$ as a more compact  \emph{explicit substitution} $\tm\esub{\var}{\tmtwo}$ (\ES for short), which binds $\var$ in $\tm$. Moreover, our $\letexp$/ES does not fix an order of evaluation between $\tm$ and $\tmtwo$, in contrast to many papers in the literature (\eg Sabry and Wadler \cite{DBLP:journals/toplas/SabryW97} or Levy et al. \cite{DBLP:journals/iandc/LevyPT03}) where $\tmtwo$ is evaluated first. Intuitively, $\tm\esub{\var}{\tmtwo}$ is a construct for \emph{sub-term sharing}, as for instance is evident when comparing $\tm\tm$ and $(\var\var)\esub\var\tm$, where $\tm$ is shared in the latter.

The set of free variables of a term $\tm$ is denoted by $\fv{\tm}$ and terms are identified up to $\alpha$-renaming. We use $\tm \isub{\var}{\tmtwo}$ for the capture-avoiding substitution of
$\tm$ for each free occurrence of $\var$ in $\tm$.

\myparagraph{Contexts} We shall use many notions of \emph{contexts}, \ie terms with a hole noted $\ctxhole$. The most general contexts used here are  open contexts $\octx$, for which the hole cannot appear under abstraction and where the adjective \emph{open} means \emph{possibly open} (that is, possibly with occurrences of free variables), and not \emph{necessarily open}. We shall also extensively use \emph{substitution contexts} $\sctx$, which are lists of \ESs ($\sctx$ stands for \emph{L}ist). Plugging a term $\tm$ in a context $\octx$ is noted $\octxp{\tm}$ and can capture variables, for instance $(\la\var\la\vartwo\ctxhole)\ctxholep{\var\vartwo} = \la\var\la\vartwo\var\vartwo$ (while $(\la\var\la\vartwo\varthree)\isub\varthree{\var\vartwo} = \la{\var'}\la{\vartwo'}\var\vartwo$); we use $\octxfp\tm$ when we want to prevent it.

Substitution contexts are in particular used to define \emph{answers}, which are abstractions surrounded by a list of substitutions. Answers shall play a key role in relating \lvsc with the positive $\l$-calculus in \refsect{translation}.
%An \emph{answer} is a term of shape $\sctxp\val$.%, where $\val$ is a value and $\sctx$ is a substitution context.

\myparagraph{Reduction Rules} The reduction rules of \VSC are slightly unusual as they use \emph{contexts} both to allow one to reduce redexes located in sub-terms, which is standard, \emph{and} to define the redexes themselves, which is less standard---this kind of rules is %sometimes 
called \emph{at a distance}. The rewriting rules at a distance of \lvsc are related to cut-elimination on proof nets, via the \cbv translation $(A \Rightarrow B)^v = \oc (A^v \multimap B^v)$ of intuitionistic logic into linear logic, see Accattoli \cite{DBLP:journals/tcs/Accattoli15}.  The linear logic connection is also the reason behind the \emph{multiplicative} and \emph{exponential} terminology for the rewriting rules. 

The multiplicative root rule $\rtom$ turns a $\beta$-redex (at a distance, because of the substitution context $\subctx$) into an ES. Note that there is no \emph{by-value} restriction, rule $\rtom$ can fire also if the argument is not a value. The by-value restriction is part of rules $\rtoe$ and $\rtogc$ which take care of the substitution process.

The VSC usually appears with a single \emph{small-step} substitution / exponential rule, where small-step refers to the fact that it is based on meta-level substitution $\tm \isub{\var}{\val}$, which replaces all the occurrences of the bound variable $\var$ at once. Here, we adopt a \emph{micro-step} presentation: the exponential rule $\rtoe$ replaces a \emph{single} variable occurrence at a time, and there is an additional garbage collection rule $\rtogc$ for removing ESs bounding variables with no occurrences. These two rules can duplicate / erase only values. Both root rules are \emph{at a distance} in that they involve a substitution context $\subctx$, which is not duplicated / erased.

The rewriting relation $\toovsc$ of \lvsc is obtained by closing the root rewriting rules $\rtom$, $\rtoe$, and $\rtogc$ by open contexts $\octx$ and by taking the union of the obtained rules.

\myparagraph{Confluence and Lack of Diamond} The VSC is confluent and the open small-step VSC even has the \emph{diamond property} (defined in \refsect{rewriting-notions}), see Accattoli and Paolini \cite{DBLP:conf/flops/AccattoliP12}. In the literature, there is no proof that the variant \lvsc used here is confluent, but this can be easily proved by adjusting the proof in \cite{DBLP:conf/flops/AccattoliP12}. We omit the details since the proof is absolutely standard. For the study in this paper, instead, it is worth pointing out that \lvsc is \emph{not} diamond, as the following diagram shows:
\begin{center}
\phantomsection$\label{lvsc-not-diamond}$
\begin{tikzpicture}[ocenter]
		\node at (0,0)[align = center](source){\normalsize $\octxfp\var \esub\var\vartwo\esub\vartwo\val$};
		\node at (source.center)[right = 190pt](source-right){\normalsize $\octxfp\var \esub\var\val\esub\vartwo\val$};
		\node at (source.center)[below = 25pt](source-down){\normalsize $\octxfp\vartwo \esub\var\vartwo\esub\vartwo\val$};
		\node at (source-right|-source-down)(target){\normalsize $\octxfp\val \esub\var\val\esub\vartwo\val$};
		\node at \med{source-down.east}{target.west}(fifthnode){\normalsize $\octxfp\vartwo \esub\var\val\esub\vartwo\val$};
		
		\draw[->](source) to node[above] {\scriptsize $\osym\esym$} (source-right);
		\draw[->](source) to node[left] {\scriptsize $\osym\esym$}(source-down);
		
		\draw[->, dotted](source-right) to node[right] {\scriptsize $\osym\esym$}(target);
		\draw[->, dotted](source-down) to node[above] {\scriptsize $\osym\esym$} (fifthnode);
		\draw[->, dotted](fifthnode) to node[above] {\scriptsize $\osym\esym$} (target);
%		\draw[<->](tarcal) to (tartam);
	\end{tikzpicture}
\end{center}
%In the small-step case, the diamond property immediately implies \emph{uniform normalization} (see \refsect{rewriting-notions}). In the micro-step case that we consider, uniform normalization holds but there is no evident proof of it. It could be proved by relating the small-step and the micro-step case, or via \cbv multi types \cite{}

\myparagraphmed{Postponement of Garbage Collection} A typical property of micro-step $\l$-calculi at a distance is the possibility of postponing garbage collection (GC). In our setting, the postponement preserves both the number of non-GC steps (which is standard) and the number of GC steps (which is not always the case) because GC steps cannot be duplicated in call-by-value weak evaluation (since only values are duplicated, but GC steps cannot take place inside values because of weakness). As it is standard, the (global) postponement of GC is obtained by iterating a local form of postponement. Notation: $\towngc \ \defeq\  \tomo \cup \toeo$.
\begin{toappendix}
\begin{proposition}[Local postponement of garbage collection\NoteProof{prop:local-postponement-gc}]
For $a \in \set{\msym, \esym}$, If $\tm \togco \Rew{\osym a}
\tmtwo$, then $\tm \Rew{\wsym a} \togco \tmtwo$.\label{prop:local-postponement-gc}
\end{proposition}
\end{toappendix}

\begin{toappendix}
\begin{proposition}[Postponement of garbage collection\NoteProof{prop:gc-postponement-vsc}]
\label{prop:gc-postponement-vsc}
If $\deriv: \tm \tow^* \tmtwo$, then there exist reduction sequences $\derivtwo: \tm
\towngc^* \tmtwop$ and $\derivthree: \tmtwop \togco^* \tmtwo$ with $\sizemo{\derivtwo} = \sizemo{\deriv}$, $\sizeeo{\derivtwo} = \sizeeo{\deriv}$, and $\size{\derivthree} = \sizegco{\deriv}$.
\end{proposition}
\end{toappendix}

\myparagraphsmall{Local Termination} As it is often the case when $\beta$-reduction is decomposed into smaller rules, every single rule of \lvsc is strongly normalizing separately (it is only together that they may diverge, namely $\towngc$ diverges on some terms).
\begin{toappendix}
\begin{proposition}[Local termination\NoteProof{prop:local-termination-lsvsc}]
\label{prop:local-termination-lsvsc}
Let $a \in \set{\msym, \esym, \gcsym}$. Relation $\Rew{\osym a}$ is strongly normalizing. Moreover, $\toeo\cup\togco$ is strongly normalizing.
\end{proposition}
\end{toappendix}

\myparagraphsmall{Renaming Chains} In \lvsc, there can be renaming chains such as $\tm\esub{\var_1}{\var_2}\esub{\var_2}{\var_3}\ldots \esub{\var_{n-1}}{\var_n}$. The issue with these chains is that some terms dynamically create longer and longer chains, slowing down the evaluation process. The simplest term on which the issue can be observed is the looping combinator $\Omega$:
\begin{center}$
\begin{array}{rllllllll}
\Omega & = & (\la\var\var\var)(\la\var\var\var) 
& \tomo &
(\varfirst\varfirst)\esub{\varfirst}{\la\var\var\var} \\
& \toeo &
((\la\var\var\var)\varfirst)\esub{\varfirst}{\la\var\var\var} 
& \tomo &
(\varsec\varsec)\esub\varsec\varfirst\esub\varfirst{\la\var\var\var} \\
& \toeo &
(\varfirst\varsec)\esub\varsec\varfirst\esub\varfirst{\la\var\var\var} 
& \toeo &
((\la\var\var\var)\varsec)\esub\varsec\varfirst\esub\varfirst{\la\var\var\var} \\
& \tomo &
(\varthird\varthird)\esub\varthird\varsec\esub\varsec\varfirst\esub\varfirst{\la\var\var\var} 
& \toeo &
(\varsec\varthird)\esub\varthird\varsec\esub\varsec\varfirst\esub\varfirst{\la\var\var\var} \\
& \toeo &
(\varfirst\varthird)\esub\varthird\varsec\esub\varsec\varfirst\esub\varfirst{\la\var\var\var} 
& \toeo &
((\la\var\var\var)\varthird)\esub\varthird\varsec\esub\varsec\varfirst\esub\varfirst{\la\var\var\var}
& \cdots
\end{array}$\end{center}
Note that after each $\tomo$ step, evaluation does a sequence of $\toeo$ steps having length equal to the number of preceeding $\tomo$ steps. Easy calculations show that the number of $\toeo$ steps is then \emph{quadratic} in the number of $\tomo$ steps, for these sequences. This quadratic overhead was first pointed out and studied by Accattoli and Sacerdoti Coen \cite{DBLP:journals/iandc/AccattoliC17}. They show that it is enough to remove variables from values (as it is done in most implementative studies, but usually without an explanation for this choice), in order to remove this issue, since evaluation then rather proceeds as follows:
\begin{center}$
\begin{array}{rllllllll}
\Omega & = & (\la\var\var\var)(\la\var\var\var) 
& \tomo &
(\varfirst\varfirst)\esub{\varfirst}{\la\var\var\var} \\
& \toeo &
((\la\var\var\var)\varfirst)\esub{\varfirst}{\la\var\var\var} 
& \tomo &
(\varsec\varsec)\esub\varsec\varfirst\esub\varfirst{\la\var\var\var} \\
& \toeo &
(\varsec\varsec)\esub\varsec{\la\var\var\var}\esub\varfirst{\la\var\var\var} 
& \toeo &
((\la\var\var\var)\varsec)\esub\varsec{\la\var\var\var}\esub\varfirst{\la\var\var\var} \\
& \tomo &
(\varthird\varthird)\esub\varthird\varsec\esub\varsec{\la\var\var\var}\esub\varfirst{\la\var\var\var} 
& \toeo &
(\varthird\varthird)\esub\varthird{\la\var\var\var}\esub\varsec{\la\var\var\var}\esub\varfirst{\la\var\var\var} \\
& \toeo &
((\la\var\var\var)\varthird)\esub\varthird{\la\var\var\var}\esub\varsec{\la\var\var\var}\esub\varfirst{\la\var\var\var}
& \tomo &
\cdots
\end{array}$\end{center}
And it is easily seen that the number of $\toeo$ steps is now \emph{linear} in the number of $\tomo$ steps. The positive $\l$-calculus of the next section shall subsume this approach, by forbidding altogether ESs containing a variable, thus also removing the ambiguity of whether variables are values or not.

%%%%%%%%%%%%%%%%%%%%%%
%\input{04-positive}
%%%%%%%%%%%%%%%%%%%%%%
\section{The (Explicit) Open Positive $\l$-Calculus}
\label{sect:positive}

%%%
\begin{figure}[t!]
\centering
\fbox{
		\begin{tabular}{c}
				$
				\begin{array}{rrll}
					\textsc{Terms }\phantom{\hspace{.3cm}}\phantom{\hspace{.3cm}} & \tm, \tmtwo, \tmthree & \grameq \var \mid \tm\esub\var{\vartwo\varthree} \mid \tm \esub\var{\la\vartwo\tmtwo} 
					\\
					\textsc{Evaluation Ctxs }\phantom{\hspace{.3cm}}\phantom{\hspace{.3cm}} & \psubctx & \grameq \ctxhole \mid \psubctx \esub\var{\vartwo\varthree} \mid \psubctx \esub\var{\la\vartwo\tm}
					
				\end{array}$
			\\[12pt]
			\hline
			%%%%%%%%%%%%%%%%%
			%%%%%%% begin weak evaluation
				$\begin{array}{c}
					\textsc{Root reduction rules}
					\\[4pt]
					\begin{array}{rll}

				    \psubctxp{\tm\esub\var{\vartwo\varthree}} \esub\vartwo{\la\varfour\psubctxtwop{\varfourp}} 
				    & \rtomep &
					\psubctxp{\psubctxtwop{\tm\isub{\var}{\varfourp}}\isub{\varfour}{\varthree}}\esub\vartwo{\la\varfour\psubctxtwop{\varfourp}} 
					\\
					\tm\esub\var{\la\vartwo\tmtwo} &  \rtogcp & \tm \ \ \ \  \mbox{if }\var\notin\fv\tm
					\end{array}
					\end{array}$
					\\[25pt]
					\hline
					\\[-15pt]
					$\begin{array}{r|c}
					\textsc{Notation} \ \ \topos    \defeq  \tomep \cup \togcp
					&
					\begin{array}{rl}
					\textsc{Ctx closure}\phantom{\hspace{.3cm}}\phantom{\hspace{.3cm}}
					&
					\RightLabel{$\asym{\in}\{\mepsym,\gcpsym\}$}
					\AxiomC{$\tm \rootRew{\asym} \tm'$}
					\UnaryInfC{$\psubctxp\tm \Rew{\wsym\asym} \psubctxp{\tm'}$}
					\DisplayProof
					\end{array}
				\end{array}$
				\end{tabular}
			}
				\caption{The open positive $\l$-calculus \Lpos.}
				\label{fig:positive}	
		\end{figure}
%%%
Here, we present the open fragment \Lopos of Wu's positive $\l$-calculus \Lpos, and then slightly refine it into an \emph{explicit} variant \Loxpos. The definition of \Lopos is in \reffig{positive}. Terms of \Lpos are a subset of VSC terms.

\myparagraph{Terms} In \Lopos, as in the $\l$-calculus, there are only three constructors, variables, applications, and abstractions. There are however, various differences, namely:
\begin{itemize}
\item Applications are only between variables, that is, of the shape $\vartwo\varthree$;
\item Applications and abstractions are always shared, that is, standalone applications and abstractions are not allowed in the grammar. They can only be introduced by the explicit substitution constructs $\esub{\var}{\vartwo\varthree}$ and
$\esub{\var}{\la\vartwo\tmtwo}$;
\item Positive sharing is peculiar as positive terms are \emph{not} shared in general, that is, $\tm\esub\var\tmtwo$ is not a positive term. There are only two distinct sharing constructors for applications and abstractions, but no construct for sharing variables or applications/abstractions with top-level sharing;
\end{itemize}

%%%
\begin{figure}[t!]
    \centering
    \fbox{
      \begin{tabular}{c}
        \RightLabel{$var$}
        \AxiomC{}
        \UnaryInfC{$\forms, \var : \typtm \vdash \var : \typtm$}
        \DisplayProof
        \qquad
%        \RightLabel{$app$}
%        \LeftLabel{$set(\vartwo : \typtmtwo\Rightarrow\typtmthree, \varthree : \typtmthree) \subseteq \forms$}
%        \AxiomC{$\forms, \var : \typtmthree \vdash \tm : \typtm$}
%        \UnaryInfC{$\forms \vdash \tm\esub\var{\vartwo\varthree} : \typtm$}
%        \DisplayProof 
        \RightLabel{$app$}
        \AxiomC{$\forms, \vartwo : \typtmtwo\Rightarrow\typtmthree, \varthree : \typtmtwo, \var : \typtmthree \vdash \tm : \typtm$}
        \UnaryInfC{$\forms, \vartwo : \typtmtwo\Rightarrow\typtmthree, \varthree : \typtmtwo \vdash \tm\esub\var{\vartwo\varthree} : \typtm$}
        \DisplayProof 
        \\[15pt]

        \RightLabel{$abs$}
        \AxiomC{$\forms, \vartwo : \typtmtwo \vdash \tmtwo : \typtmthree$}
        \AxiomC{$\forms, \var : \typtmtwo\Rightarrow\typtmthree \vdash \tm : \typtm$}
        \BinaryInfC{$\forms \vdash \tm\esub\var{\la\vartwo\tmtwo} : \typtm$}
        \DisplayProof
      \end{tabular}
        }
                    \caption{Assignment of simple types to positive $\lambda$-terms.}
                    \label{fig:simple-types}	
            \end{figure}
%%%
\myparagraph{Types} We are not going to work with types, and yet in \reffig{simple-types} we show how simple types can be assigned to positive terms. The main point is to suggest that the \emph{positive focusing} aspect, roughly, amounts to have only rules that act on the \emph{left} of the deduction symbol $\vdash$, leaving the right hand type $\typtm$ unchanged. Note that this is dual to what happens in the standard system of simple types for the $\l$-calculus (\ie natural deduction), which correspond to the negative polarity assignment, where rules only act on the \emph{right} of $\vdash$.

\myparagraph{Contexts} The notions of substitution contexts $\sctx$ and open contexts $\octx$ of the \lvsc pleasantly collapse onto a single notion of \emph{evaluation contexts} $\psubctx$ in \Lopos, because the additional clauses for $\octx$ in \lvsc (for the sub-terms of applications and inside ESs) do not make sense in \Lopos, due to the restricted shape of terms. As it is immediately seen from the grammar of positive terms, every positive term $\tm$ can be written uniquely as $\psubctxp\var$ for some $\var$ and $\psubctx$, with $\psubctx$ possibly capturing $\var$.

\myparagraph{Rewriting Rules} There are two rewriting rules at a distance, based on open contexts and defined in \reffig{positive}. Rule $\togcp$ handles garbage collection and is standard. Rule $\tomep$ is quite heavy, essentially it is a macro reduction step concatenating two steps of the VSC plus a meta-level substitution. Let us explain it (it is not necessary to grasp it completely, since right after we shall introduce a slight variant of the positive calculus that has simpler rewriting rules). Rule $\tomep$ does three operations, and also adopts some notations to respect the constrained shape of terms:
\begin{enumerate}
\item \emph{Useful exponential step}: it does the useful\footnote{Useful steps shall be defined and studied in the next section. Here we rest on the intuitive description given in the introduction.} replacement of an applied occurrence of $\vartwo$, i.e. it acts on $\vartwo$ on $\psubctxp{\tm\esub\var{\vartwo\varthree}} $, by an abstraction $\la\varfour\tmtwo$. 
\item \emph{Created multiplicative step}: since $\tm\esub\var{(\la\varfour\tmtwo)\varthree}$ is not a construct of \Lpos, the rule has to also do on-the-fly what in the VSC is a multiplicative step, which would create the ES $\tmtwo\esub\varfour\varthree$. 
\item \emph{Further meta-level substitution}: but since $\tmtwo\esub\varfour\varthree$ is not a construct of \Lpos either, the step also does on-the-fly the meta-level substitution associated to $\esub\varfour\varthree$. 
\item \emph{Respecting the syntax}: to write the reduct, one needs to respect the constrained shape of positive terms, which requires to write $\tmtwo$ as $\psubctxtwop{\varfourp}$ and then do the re-arrangement of the term structure and the meta-level substitutions of variables specified by the rule in \reffig{positive}.
\end{enumerate}
Working with such a rule is heavy. A first reason is that, quite simply, it involves a lot of symbols. Another reason is that, by concatenating steps of different nature of the VSC, \Lopos is \emph{not} conservative over \lvsc in the sense that it breaks one of its key property, namely local termination (stated by \refprop{local-termination-lsvsc}). Rule $\tomep$ can indeed diverge by itself, as in the following representation of the looping term $\Omega$ in \Lopos:
\begin{center}$\begin{array}{llllllll}
  \var\esub{\var}{\vartwo\vartwo}\esub{\vartwo}{\la\varthree{\varfour\esub{\varfour}{\varthree\varthree}}}
&  \tomep&
  \var\esub{\var}{\vartwo\vartwo}\esub{\vartwo}{\la\varthree{\varfour\esub{\varfour}{\varthree\varthree}}}
&  \tomep&
  \cdots
  \end{array}
$\end{center}
We are then going to decompose $\tomep$ into two rules, by adding a new constructor to the calculus.

%%%
\begin{figure}[t!]
\centering
\fbox{
		\begin{tabular}{c}
				$
				\begin{array}{rrll}
		\textsc{Terms } & \tm, \tmtwo, \tmthree & \grameq \var \mid \tm \esub\var{\vartwo\varthree} \mid \tm \esub\var{\la\vartwo\tmtwo} \mid \esub{\var}{(\la\vartwo\tmtwo)\varthree}
					\\
					\textsc{Evaluation Ctxs } & \ppsubctx & \grameq \ctxhole \mid \ppsubctx\esub{\var}{\vartwo\varthree} \mid \ppsubctx\esub{\var}{\la\vartwo\tm} \mid \ppsubctx\esub{\var}{(\la\vartwo\tm)\varthree} 					
				\end{array}$
			\\[13pt]
			\hline
			%%%%%%%%%%%%%%%%%
			%%%%%%% begin weak evaluation
				$\begin{array}{c}
					\textsc{Root reduction rules}
					\\[4pt]
					\begin{array}{rlll}
  				    \tm\esub\var{(\la\vartwo{\ppsubctxp{\varthree}})\varfour}
				    & \rtomp &
\ppsubctxp{\tm\isub{\var}{\varthree}}\isub{\vartwo}{\varfour}
					\\
				    \ppctxp{\tm\esub\var{\vartwo\varthree}}
\esub\vartwo{\la\varfour\tmtwo}
				    & \rtoep &
\ppctxp{\tm\esub{\var}{(\la\varfour\tmtwo)\varthree}}\esub\vartwo{\la\varfour\tmtwo}
					\\
					\tm\esub\var{\la\vartwo\tmtwo} &  \rtogcp & \tm \ \ \ \  \mbox{if }\var\notin\fv\tm
					\end{array}
					\end{array}$
			\\[33pt]
					\hline
			\\[-15pt]
					$\begin{array}{r|c}
					\textsc{Notation} \ \ 					\toepos    \defeq  \tomp \cup \toep\cup \togcp
					&
					\begin{array}{rl}
					\textsc{Ctx closure}
					&
					\RightLabel{$\asym{\in}\{\mpsym,\epsym,\gcpsym\}$}
					\AxiomC{$\tm \rootRew{\asym} \tm'$}
					\UnaryInfC{$\ppsubctxp\tm \Rew{\wsym\asym} \ppsubctxp{\tm'}$}
					\DisplayProof
					\end{array}
				\end{array}$
				\end{tabular}
			}
				\caption{The open explicit positive $\l$-calculus \Lppos.}
				\label{fig:explicit-positive}	
		\end{figure}
%%%
\myparagraph{The Explicit Positive $\l$-Calculus} The explicit positive calculus \Lppos is defined in \reffig{explicit-positive}. The idea is to extend \Lopos by adding the intermediate construct $\tm\esub\var{(\la\varfour\tmtwo)\varthree}$ which allows us to decompose rule $\rtomep$ in two, separating the first operation, the useful exponential step, now noted $\rtoep$, from the rest of the rule, now noted $\rtomp$. Rule $\rtomp$ still corresponds to two actions (the multiplicative step of the VSC and the meta-level substitution of the variable), but it is one of the key points of \Lopos that variable substitutions disappear, so we shall not decompose the rule further. Intuitively, the new construct $\tm\esub\var{(\la\varfour\tmtwo)\varthree}$ plays a similar (but dual) role to the introduction of ESs in the $\l$-calculus to decompose $\beta$-redexes in two, as it is an explicit $\beta$-redex. This is why we refer to the obtained calculus as to the \emph{explicit positive calculus}. One might also see it as switching the shape of (compact) applications of \Lpos from $\var\vartwo$ to $\val\vartwo$.

Clearly, \Lppos simulates \Lopos: if $\tm\tomep\tmtwo$ then $\tm\toep\tomp\tmtwo$.
\begin{proposition}
\Lppos simulates \Lopos.
\end{proposition}

\myparagraphmed{Diamond} A difference between \lvsc and \Lppos is that the latter is diamond, as we now show, while the former is not (see \refsect{vsc}). The proof uses the following basic lemma.
\begin{lemma}[Stability under renamings]
\label{l:stability-renamings}
Let $\tm$ and $\tmtwo$ be \Lppos terms. 
If $\tm \toepos \tmtwo$ then $\tm\isub\var\vartwo \toepos \tmtwo\isub\var\vartwo$ for any $\var$ and $\vartwo$.
\end{lemma}

\begin{toappendix}
\begin{theorem}[Positive diamond\NoteProof{thm:lppos-diamond}]
Relation $\toepos$ is diamond.\label{thm:lppos-diamond}
\end{theorem}
\end{toappendix}

\myparagraphmed{Postponement of GC and Local Termination} In terms of properties, \Lppos is conservative with respect to \lvsc, as postponement of GC and local termination still hold. Notation: $\toepngc \defeq \tomp \cup \toep$.

\begin{toappendix}
\begin{proposition}[Local postponement of garbage collection\NoteProof{prop:local-postponement-gc-xpos}]
\label{prop:local-postponement-gc-xpos}
Let $\tm$ and $\tmtwo$ be \Lppos terms and $a \in \set{\posmsym, \posesym}$. If $\tm \togcp \Rew{\wsym a} \tmtwo$, then $\tm \Rew{\wsym a} \togcp \tmtwo$.
\end{proposition}
\end{toappendix}

\begin{toappendix}
\begin{proposition}[Postponement of garbage collection\NoteProof{prop:vsc-gc-postponement}]
\label{prop:vsc-gc-postponement}
Let $\tm$ and $\tmtwo$ be \Lppos terms, $\deriv: \tm \toepos^* \tmtwo$. Then there exist reduction sequences $\derivtwo:
 \tm \toepngc^* \tmtwop$ and $\derivthree: \tmtwop \togcp^* \tmtwo$ with
$\sizemp{\derivtwo} = \sizemp{\deriv}$, $\sizeep{\derivtwo} = \sizeep{\deriv}$,
and $\size{\derivthree} = \sizegcp{\deriv}$.
\end{proposition}
\end{toappendix}

\begin{toappendix}
\begin{proposition}[Local termination\NoteProof{prop:local-termination-positive}]
Let $a \in \{\mpsym,\epsym,\gcpsym\}$. Relation $\Rew{\osym a}$ is strongly normalizing. Moreover, $\toep\cup\togcp$ is strongly normalizing.\label{prop:local-termination-positive}
\end{proposition}
\end{toappendix}

\myparagraphmed{Absence of Renaming Chains} At the end of \refsect{vsc}, we discussed renaming chains and their dynamic creations by the looping term $\Omega$. The following term is the representation of $\Omega$ in \Lppos, together with its evaluation, where evidently the number of $\toep$ steps is linear in the number of $\tomp$ steps:
\begin{center}$
\begin{array}{rllllllllll}
&\varfour\esub\varfour{(\la\var\vartwo\esub\vartwo{\var\var})\varthree}\esub\varthree{\la\var\vartwo\esub\vartwo{\var\var}}
& \tomp &
\varfour\esub\varfour{\varthree\varthree}\esub\varthree{\la\var\vartwo\esub\vartwo{\var\var}}
\\
 \toep &
\varfour\esub\varfour{(\la\var\vartwo\esub\vartwo{\var\var})\varthree}\esub\varthree{\la\var\vartwo\esub\vartwo{\var\var}}
& \tomp &
\varfour\esub\varfour{\varthree\varthree}\esub\varthree{\la\var\vartwo\esub\vartwo{\var\var}}
\\
 \toep &
\varfour\esub\varfour{(\la\var\vartwo\esub\vartwo{\var\var})\varthree}\esub\varthree{\la\var\vartwo\esub\vartwo{\var\var}}
& \cdots &
\end{array}$
\end{center}

%%%%%%%%%%%%%%%%%%%%%%
%\input{05-dissecting}
%%%%%%%%%%%%%%%%%%%%%%
\section{Dissecting \lvsc: Variable and Useful Steps}
\label{sect:dissecting}

In this section, we isolate various sub-reductions of \lvsc in order to relate \lvsc and \Lppos in the next section. Essentially, some steps of \lvsc cannot be expressed in \Lppos, and some are instead \emph{absorbed}, that is, mapped to identities rather than being simulated. We then have to partition the rewriting rules of \lvsc into sub-rules as to identify the steps that cannot be expressed, those that are absorbed, and those that are simulated by \Lppos. In particular, the partition of steps leads us to discuss useful sharing.

%%%
\begin{figure}[t!]
\centering
\fbox{
\begin{tabular}{c}
$\begin{array}{lrcl}
  \textsc{Exp. root rule for abstractions} &\phantom{\hspace{.3cm}}\phantom{\hspace{.3cm}} \octxfp{\var}\esub\var{\subctxp{\la\vartwo\tm}} &
\rtoems & \subctxp{\octxfp{\la\vartwo\tm}\esub{\var}{\la\vartwo\tm}}
\\
  \textsc{Exp. root rule for variables} &
\octxfp{\var}\esub{\var}{\subctxp{\vartwo}} & \rtoevar &
\subctxp{\octxfp{\vartwo}\esub{\var}{\vartwo}}
\end{array}$
\\[12pt]\hline
\\[-17pt]
$\begin{array}{lrcll}
  \textsc{GC root rule for abstractions}\phantom{\hspace{.3cm}}\phantom{\hspace{.3cm}} &\phantom{\hspace{.3cm}}\phantom{\hspace{.3cm}}\phantom{\hspace{.3cm}} \tm\esub\var{\subctxp{\la\vartwo\tmtwo}} &
\rtogcabs & \subctxp{\tm} & \mbox{if }\var\notin\fv\tm
\\
  \textsc{GC root rule for variables} &
\tm\esub{\var}{\subctxp{\vartwo}} & \rtogcvar &
\subctxp\tm & \mbox{if }\var\notin\fv\tm
\end{array}$
\\[12pt]\hline
\\[-15pt]
					$\begin{array}{lr}
					\textsc{Ctx closure}\phantom{\hspace{.3cm}}\phantom{\hspace{.3cm}}\phantom{\hspace{.3cm}}\phantom{\hspace{.3cm}}\phantom{\hspace{.3cm}}\phantom{\hspace{.3cm}}\phantom{\hspace{.3cm}}\phantom{\hspace{.3cm}}
					&\phantom{\hspace{.3cm}}\phantom{\hspace{.3cm}}\phantom{\hspace{.3cm}}\phantom{\hspace{.3cm}}\phantom{\hspace{.3cm}}\phantom{\hspace{.3cm}}\phantom{\hspace{.3cm}}
					\RightLabel{$\asym{\in}\{\eabssym,\evarsym,\gcabssym, \gcvarsym\}$}
					\AxiomC{$\tm \rootRew{\asym} \tm'$}
					\UnaryInfC{$\ppsubctxp\tm \Rew{\wsym\asym} \ppsubctxp{\tm'}$}
					\DisplayProof
					\end{array}$
\end{tabular}
}
\caption{Dissected rewriting rules for \lvsc.}
\label{fig:dissecting}
\end{figure}
%%%
\myparagraph{Variable Substitutions} In \Lppos, there is no way to represent an ES $\tm\esub\var\vartwo$ containing a variable, and there is also no way of simulating a variable exponential step such as $\octxfp\var\esub\var\vartwo \toe \octxfp\vartwo\esub\var\vartwo$ or a variable GC step $\tm\esub\var\vartwo \toe \tm$ with $\var\notin\fv\tm$. These steps shall be \emph{absorbed} by our translation from \lvsc to \Lppos, that is, they shall be mapped to identities. To properly state this fact later on, we split now the root exponential rule $\rtoe$ in two rules $\rtoems$ and $\rtoevar$, depending on whether the replacing value is an abstraction or a variable, and similarly for GC. The split rules are defined in \reffig{dissecting}.

Of two sub-rules $\rtoems$ and $\rtogcabs$ that are not absorbed, $\rtogcabs$ shall be closed by all open context and simply factored-out via the postponement of GC (\refprop{gc-postponement-vsc}). %Thus, for the sake of simplicity, we shall not simulate it in \Lppos. 
The other sub-rule $\rtoems$ is where usefulness plays a role, discussed next.

\myparagraph{Useful Steps, First Intuitions}  Another difference between the two calculi is that in \Lppos exponential steps can only be directly \emph{useful}. Let us overview usefulness in a bit more detail than in the introduction. Replacements of variable occurrences out of ESs by abstractions amount to three \emph{direct} cases, in the \lvsc:
\begin{equation}
\arraycolsep=3pt
\begin{array}{r rll}
\textsc{Directly useful} &\phantom{\colspace} (\var\tm)\esub\var{\la\vartwo\tmtwo} & \rtoems &  ((\la\vartwo\tmtwo)\tm)\esub\var{\la\vartwo\tmtwo}
\\
\textsc{Directly non-useful 1} &\phantom{\colspace} (\tm\var)\esub\var{\la\vartwo\tmtwo} & \rtoeabs &  (\tm (\la\vartwo\tmtwo))\esub\var{\la\vartwo\tmtwo}
\\
\textsc{Directly non-useful 2} &\phantom{\colspace} \var\esub\var{\la\vartwo\tmtwo} & \rtoeabs &  (\la\vartwo\tmtwo)\esub\var{\la\vartwo\tmtwo}
\end{array}
\label{eq:direct-usef-body}
\end{equation}
The terminology \emph{useful} refers to the fact that the exponential step \emph{creates} a new multiplicative redex, namely $(\la\vartwo\tmtwo)\tm$, and it is used to contrast with \emph{non-useful}, or \emph{useless} exponential steps (we shall prefer \emph{non-useful} in this paper, as to denote them concisely with the letter \emph{n}, given that \emph{useful} and \emph{useless} both start with \emph{u}) that instead do not create multiplicative steps. 

The reason why one considers usefulness with respect to multiplicative steps is that the number of multiplicative steps is a reasonable cost model in \cbv \cite{DBLP:conf/lics/AccattoliC15,DBLP:journals/scp/AccattoliG19,DBLP:conf/lics/AccattoliCC21}, and that this fact is proved by an evaluation strategy that crucially avoids non-useful substitution steps, since non-useful substitution steps can at times add an exponential overhead, breaking the reasonability of the cost model.

In \Lppos, directly useful steps can be simulated, while the two kinds of direct non-useful step cannot be expressed. The first kind because of the shape of positive applications, which can only have a variable as an argument. The second kind because abstractions cannot appear out of ESs in \Lppos (this is the relevance of the second aspect of the compactness of \Lpos/\Lppos mentioned in the introduction).

There are (at least) two technical difficulties in defining the useful steps of \lvsc precisely. Before diving into them, we provide a \emph{disclaimer}. Useful sharing is complex and takes different shapes in different settings (i.e. call-by-name/value/need). For this reason, the only two works distinguishing into useful and non-useful steps, one in call-by-name \cite{DBLP:journals/corr/AccattoliL16} and one in \cbv \cite{DBLP:conf/lics/AccattoliC15}, rest on slightly different definitions concerning indirectly (non-)useful steps (defined below). On purpose, later papers avoid these definitions, despite implementing useful sharing \cite{DBLP:journals/scp/AccattoliG19,DBLP:conf/lics/AccattoliCC21,DBLP:conf/csl/AccattoliL22}. Here, we are going to define the two kinds of steps, but our definitions (and terminology) shall\textemdash once more\textemdash be slightly different from \cite{DBLP:journals/corr/AccattoliL16,DBLP:conf/lics/AccattoliC15}. The only thing on which all these works agree are the notions of direct useful steps, which is what \Lppos captures.

\myparagraph{Difficulty 1: Indirect Useful Steps} In \lvsc, there are also \emph{indirect} cases to consider, given by when there is a renaming chain connecting the acting ESs and the end variable occurrence. Consider for instance the following three indirect cases, mimicking the direct ones above via a renaming chain of length 1:
\begin{equation}
\arraycolsep=3pt
\begin{array}{r rll}
\textsc{Indirectly useful} &\phantom{\colspace} (\var\tm)\esub\var\varthree\esub\varthree{\la\vartwo\tmtwo} & \rtoems &  (\var\tm)\esub\var{\la\vartwo\tmtwo}\esub\varthree{\la\vartwo\tmtwo}
\\
\textsc{Indirectly non-useful 1} &\phantom{\colspace} (\tm\var)\esub\var\varthree\esub\varthree{\la\vartwo\tmtwo} & \rtoeabs &  (\tm\var)\esub\var {\la\vartwo\tmtwo}\esub\varthree{\la\vartwo\tmtwo}
\\
\textsc{Indirectly non-useful 2} &\phantom{\colspace} \var\esub\var\varthree\esub\varthree{\la\vartwo\tmtwo} & \rtoeabs &  \var\esub\var {\la\vartwo\tmtwo}\esub\varthree{\la\vartwo\tmtwo}
\end{array}\label{eq:indirect-usef-body}
\end{equation}
The first step does not create a multiplicative redex, but it is usually considered as useful because it contributes anyway to the \emph{future} creation of the multiplicative redex that shall happen after that step. Similarly, the other two cases are usually considered as non-useful. This indirect aspect is quite difficult to work with.  In \Lppos, the indirect phenomenon disappears, because ESs such as $\esub\var\varthree$, which are the cause of the indirection, do not exist. This is a considerable improvement. 

It remains, however, the issue of relating indirectly useful steps in \lvsc with (directly useful) steps in \Lppos. We somewhat circumvent this issue by departing from the literature and considering indirect useful steps \emph{as non-useful}. Therefore, \emph{all the indirect cases above are non-useful}, in this paper. This is not cheating, as we shall explain, it is related to the core factorization theorem of the next section. 

For the sake of completeness, one should also consider the replacements in which the end variable is inside an ES. The first two cases of both \refeq{direct-usef-body} and \refeq{indirect-usef-body} are unaffected. The third one instead amounts to extend a chain, and then can become any of the three cases in \refeq{indirect-usef-body}. 
%Another related point in which we depart from the literature is that we do not assign a useful/non-useful character to $\toevar$ steps. The idea is that given

\myparagraph{Difficulty 2: Contextual Closure} The other difficulty is that the directly useful steps of \lvsc cannot be defined at the root level and then closed by evaluation contexts, since sometimes the useful aspect is contributed by the evaluation context itself. Consider the following root step:
\begin{center}
$\begin{array}{ccc}
\var\esub\var{\la\vartwo\tmtwo} \ \ \rtoems\ \  (\la\vartwo\tmtwo)\esub\var{\la\vartwo\tmtwo}
\end{array}$
\end{center}
As a root step it is not useful. But when plugged in the evaluation context $\octx=\ctxhole\tm$, it gives rise to the following useful step, since now the steps creates a multiplicative redex in the reduct, because of the definition at a distance of these redexes:
\begin{center}
$\begin{array}{ccc}
\var\esub\var{\la\vartwo\tmtwo}\tm \ \ \rtoems\ \  (\la\vartwo\tmtwo)\esub\var{\la\vartwo\tmtwo}\tm
\end{array}$
\end{center}
Therefore, usefulness of a $\toeabs$ step depends also on the evaluation contexts surrounding the root step.

\myparagraph{Useful Contexts and Steps} Useful exponential steps are defined via useful contexts by putting together the context used to isolate the replaced variable and the surrounding evaluation context. We also define non-useful contexts $\badctx$, whose first two clauses cover the two direct cases in \refeq{direct-usef-body} and whose third clause cover the three indirect cases in \refeq{indirect-usef-body}.
\begin{definition}
Useful and non-useful contexts of \lvsc are defined as follows:
\begin{center}
$\arraycolsep=3pt\begin{array}{r rll r rll}
\textsc{Useful ctxs} &\phantom{\hspace{.3cm}} \goodctx & \grameq &\octxp{\subctx\tm} &\phantom{\hspace{.3cm}}\phantom{\hspace{.3cm}}\phantom{\hspace{.3cm}}\phantom{\hspace{.3cm}}\phantom{\hspace{.3cm}}
\textsc{Non-useful ctxs} &\phantom{\hspace{.3cm}} \badctx & \grameq &\subctx \mid \octxp{\tm\subctx} \mid \octxp{\tm\esub{\var}{\subctx}}
\end{array}$
\end{center}
When taking into account the evaluation context, an $\toeabs$ step has the following shape:
\begin{center}
$\octxfirstp{\octxsec\ctxholefp{\var}\esub{\var}{\subctxp{\la\vartwo\tm}}} \toeabs
\octxfirstp{\subctxp{\octxsec\ctxholefp{\la\vartwo\tm}\esub{\var}{\la\vartwo\tm}}}$
\end{center}
Such a $\toeabs$ step is \emph{useful} if $\octxfirstp{\octxsec\esub{\var}{\subctxp{\la\vartwo\tm}}}$ is a useful context, and \emph{non-useful} otherwise.
\end{definition}

\myparagraphmed{Properties of Useful Contexts} In symbols, we use predicates $\goodpred$ and $\badpred$: $\isgood{\octx}$ means that $\octx$ is \good while $\isbad{\octx}$ means that $\octx$ is \bad. Similarly, we use the predicate $\subpred$ (resp. $\notsubpred$) for substitution contexts (resp. non-substitution contexts).

The following immediate lemma states how useful contexts depend on their sub-contexts, stressing that the only subtle case is the first one.
\begin{lemma}[Useful sub-contexts] % \reflemmap{goodctx}{appl}
\label{l:goodctx}
\hfill
\begin{enumerate} 
  \item \label{p:goodctx-appl} $\isgood{\octx\tm} \Leftrightarrow \isgood{\octx}
\vee \issub{\octx}$.
  \item \label{p:goodctx-appr} $\isgood{\tm\octx} \Leftrightarrow \isgood{\octx}$.
  \item \label{p:goodctx-subctx} $\isgood{\octx\esub{\var}{\tm}} \Leftrightarrow
\isgood{\octx}$.
  \item \label{p:goodctx-es} $\isgood{\tm\esub{\var}{\octx}} \Leftrightarrow
\isgood{\octx}$.
\end{enumerate}
\end{lemma}

As a sanity check, we show that the definitions of \good and \bad contexts provide a partition of open contexts. The proof is an easy induction on the open context $\octx$, the only non immediate case of which  is the one for $\octx=\octxtwo \tm$, as it can be guessed by \reflemma{goodctx}.
\begin{toappendix}
\begin{lemma}[Useful partitions open\NoteProof{l:good-bad}]
\label{l:good-bad}
A VSC open context $\octx$ is either \good or \bad.
\end{lemma}
\end{toappendix}
%%%
\begin{figure}
\centering
\fbox{
$\begin{array}{r r@{\ }l@{\ }l}
\textsc{\Good exp. rule} &\phantom{\hspace{.3cm}}
\octxfirstp{\octxsec\ctxholefp{\var}\esub\var{\subctxp{\la\vartwo\tm}}} &
\toegood &
\octxfirstp{\subctxp{\octxsec\ctxholefp{\la\vartwo\tm}\esub{\var}{\la\vartwo\tm}}} \text{ if }
\isgood{\octxfirstp{\octxsec}} \\
\textsc{\Bad exp. rule} &
\octxfirstp{\octxsec\ctxholefp{\var}\esub\var{\subctxp{\la\vartwo\tm}}} &
\toebad & \octxfirstp{\subctxp{\octxsec\ctxholefp{\la\vartwo\tm}\esub{\var}{\la\vartwo\tm}}} \text{ if }
\isbad{\octxfirstp{\octxsec}} \\
\end{array}$
}
\caption{Definition of \good and \bad exponential variants of $\toeabs$, based on \reflemma{good-ctx-properties}.1.}
\label{fig:dissecting-eabs}
\end{figure}
%%%
The next easy lemma collects a few properties that are helpful in proofs. The first point says that in a $\toeabs$ step the usefulness of the context does not depend on the acting substitution $\esub{\var}{\subctxp{\la\vartwo\tm}}$ but only on the composition of the inner and outer open contexts $\octxfirst$ and $\octxsec$. The definition of (non-)useful steps is then re-stated in \reffig{dissecting-eabs}.
\begin{toappendix}
\begin{lemma}[Context plugging and usefulness\NoteProof{l:good-ctx-properties}]
\label{l:good-ctx-properties}\hfill
\begin{enumerate}
\item 
%\label{l:goodctx-plugging-subctx}
$\isgood{\octxfirstp{\subctxp{\octxsec}}} \Leftrightarrow
\isgood{\octxfirstp{\octxsec}}$.
\item %\label{l:goodctx-plugging}
$\isgood{\octxfirstp{\octxsec}} \Leftrightarrow \isgood{\octxsec} \vee
(\issub{\octxsec} \wedge \isgood{\octxfirst})$.
\item %\label{l:badctx-plugging}
$\isbad{\octxfirstp{\octxsec}} \Leftrightarrow \isbad{\octxsec} \wedge
(\issub{\octxsec} \Rightarrow \isbad{\octxfirst}$.
\end{enumerate}
\end{lemma}
\end{toappendix}

%%%%%%%%%%%%%%%%%%%%%%
%\input{06-core_factorization}
%%%%%%%%%%%%%%%%%%%%%%
\section{Core Factorization, or Postponing Non-Useful Steps}
\label{sect:core-factorization}
Since \bad steps cannot be simulated by the positive calculus, the translation from  \lvsc to \Lppos of the next section cannot induce a bisimulation---we need a finer approach. The idea is that any reduction sequence in \lvsc can be factored into a core part (that includes useful steps, that is, $\toegood$) and a non-useful part, and that the evaluation of a term terminates if and only if its core evaluation terminates. In this section, we prove these two facts. In the next one, we shall give a translation from  \lvsc and \Lppos inducing a simulation between the core part of \lvsc and \Lppos, and a termination equivalence result.

\myparagraph{Core Reduction} The (open) core reduction $\tocore$ of \lvsc is defined as $\tocore \defeq \tomo \cup \toegood \cup \toevar$, and we dub \emph{Core \lvsc} the set of VSC terms endowed with $\tocore$. Beyond multiplicative and useful exponential steps, which shall be simulated by \Lppos, it includes also $\toevar$ steps, which\textemdash on purpose\textemdash we have \emph{not} classified as useful or non-useful (in another departure from the literature) and which are going to be absorbed. As we explain next, they are crucial for our core factorization theorem.

\myparagraph{Non-Useful Postponement} Factorization can be seen as a postponement property, in this case of \bad steps. Similarly to how we dealt with GC, we first give a local form of postponement of $\toebad$ steps. The proof of local postponement is simple but more involved than for GC steps, since it requires to check all the (many!) possible cases for a core step following a $\toebad$ step, which are quite technical to list given the many contexts involved in the definition of (non-)useful rewriting steps. There is also a case of tricky local postponement diagram, where the swap postponing $\toebad$ requires to do \emph{two} core steps (this case is taken into account in \refprop{local-postponement-bad}.(ii) below):
\begin{center}
\begin{tikzpicture}[ocenter]
		\node at (0,0)[align = center](source){\normalsize $(\var\tm) \esub\var\vartwo\esub\vartwo{\la\varthree\tmtwo}$};
		\node at (source.center)[right = 240pt](source-right){\normalsize $(\var\tm) \esub\var{\la\varthree\tmtwo}\esub\vartwo{\la\varthree\tmtwo}$};
		\node at (source.center)[below = 25pt](source-down){\normalsize $(\vartwo\tm) \esub\var\vartwo\esub\vartwo{\la\varthree\tmtwo}$};
		\node at (source-right|-source-down)(target){\normalsize $((\la\varthree\tmtwo)\tm) \esub\var{\la\varthree\tmtwo}\esub\vartwo{\la\varthree\tmtwo}$};
		\node at \med{source-down.east}{target.west}[anchor=center](fifthnode){\normalsize $((\la\varthree\tmtwo)\tm) \esub\var\vartwo\esub\vartwo{\la\varthree\tmtwo}$};
		
		\draw[->](source) to node[above] {\scriptsize $\osym\ebadsym$} (source-right);
		\draw[->, dotted](source) to node[left] {\scriptsize $\osym\evarsym$}(source-down);
		
		\draw[->](source-right) to node[right] {\scriptsize $\osym\egoodsym$}(target);
		\draw[->, dotted](source-down) to node[above] {\scriptsize $\osym\egoodsym$} (fifthnode);
		\draw[->, dotted](fifthnode) to node[above] {\scriptsize $\osym\ebadsym$} (target);
%		\draw[<->](tarcal) to (tartam);
	\end{tikzpicture}
\end{center}
 This diagram actually justifies both our avoidance of indirect useful steps in the \lvsc and the fact that $\toevar$ steps are not labeled as useful/non-useful. Note that the solid lines are exactly what would usually be an indirectly useful step followed by a directly useful one, and that for us they are instead a non-useful step followed by a (directly) useful one. The point is that the sequence can always be re-arranged as shown by the diagram, using $\toevar$ as to turn the replacement on $\esub\var\vartwo$ into a directly non-useful one.

\begin{toappendix}
\begin{proposition}[Local postponement of $\toebad$\NoteProof{prop:local-postponement-bad}]
\label{prop:local-postponement-bad}
Let $\tm$ and $\tmtwo$ be \VSC terms. If $\tm\toebad\tocore\tmtwo$ then $\tm\tocore\toebad\tmtwo$ or $\tm\tocore\tocore\toebad\tmtwo$. More precisely:
\begin{enumerate}
\item $\toebad \tomo \ \subseteq \ \tomo \toebad $;
\item $\toebad \toegood \ \subseteq \ \toegood \toebad\cup \toevar\toegood \toebad$;
\item $\toebad \toevar \ \subseteq \ \toevar \toebad$.
\end{enumerate}
\end{proposition}
\end{toappendix}

Obtaining the global postponement property from the local one is easy because, although the number of core steps can grow with  local swaps, the number of \bad steps is preserved, and can be easily exploited for the induction lifting the local diagrams to the global property.
\begin{toappendix}
\begin{theorem}[Core Factorization / Postponement of \bad steps\NoteProof{thm:postpone-bad}]
\label{thm:postpone-bad}
Let $\tm$ and $\tmtwo$ be \VSC terms. If $\deriv: \tm \towngc^* \tmtwo$, then $\derivtwo: \tm \togoodvscn \toebad^* \tmtwo$ with $\sizemo{\derivtwo} = \sizemo{\deriv}$.
\end{theorem}
\end{toappendix}

Lastly, we use the postponement property to prove that the core sub-system of \lvsc is termination-equivalent to the whole of \lvsc, justifying the core terminology.
\begin{toappendix}
\begin{theorem}[Termination equivalence of \lvsc and Core \lvsc\NoteProof{thm:core-SN-equiv}]
%Let $\tm$ be a \VSC term.
 \label{thm:core-SN-equiv}\hfill
\begin{enumerate}
\item $\tm$ has a diverging $\toovsc$ sequence if and only if $\tm$ has a diverging $\tocore$ sequence;
\item $\tm$ is $\toovsc$-weakly normalizing if and only if $\tm$ is $\tocore$-weakly normalizing.
\end{enumerate}
\end{theorem}
\end{toappendix}

%%%%%%%%%%%%%%%%%%%%%%
%\input{07-translation}
%%%%%%%%%%%%%%%%%%%%%%
\section{Translating \lvsc to \Lppos and Simulating Core Steps}
\label{sect:translation}
In this section, we define a translation $\vtopptm{\cdot}$ from \lvsc to \Lppos and show that it induces a simulation of the core reduction $\tocore$ of \lvsc by \Lppos. There are various delicate points, concerning both the definition and the simulation, discussed all along this section.

\myparagraph{Subtlety 1: Absorption of Variables} Since ESs in \lvsc  can contain variables (as e.g. in $\tm\esub\var\vartwo$) but ESs in \Lppos cannot, the translation $\vtopptm{\cdot}$ that we shall define turns these ESs of \lvsc into meta-level substitutions of \Lppos. For instance, we shall have $\vtopptm{\tm\esub\var\vartwo} = \vtopptm\tm\isub\var\vartwo$. 
%%%
\begin{figure}[t!]
\centering
     \arraycolsep=3pt
     \fbox{
\begin{tabular}{c}
\textsc{Translation of substitution contexts}
\\[2pt]
$\begin{array}{lll lllllll}
\vtoppctx{\ctxhole} &\defeq &(\ctxhole, \isubempty)
&\phantom{\hspace{.3cm}}\phantom{\hspace{.3cm}}
   \vtoppctx{\subctx\esub{\var}{\tm}} &\defeq& (\ppsubctxtwop{\ppsubctx\isub\var\vartwo},
   \isubs\isub{\var}{\vartwo})
   &
   \text{ where } \vtoppctx{\subctx} = (\ppsubctx, \isubs)
   \text{ and } \vtopptm{\tm} = \ppsubctxtwop{\vartwo}
   \end{array}$
			\\[5pt]
			\hline

\textsc{Translation of terms}
\\
$\begin{array}{rll l}
\vtopptm{\var} & \defeq &\var
\\
\vtopptm{\la\var\tm} & \defeq & \vartwo\esub{\vartwo}{\la\var\vtopptm{\tm}}
\\
\vtopptm{\tm\esub{\var}{\tmtwo}} & \defeq &
\ppsubctxp{\vtopptm{\tm}\isub{\var}{\vartwo}} &\phantom{\hspace{.3cm}} \text{ where } \vtopptm{\tmtwo} =
\ppsubctxp{\vartwo} 
\\
\vtopptm{\subctxp{\la\var\tm}\tmtwo} & \defeq
&\ppsubctxp{\ppsubctxtwop{\vartwo\esub{\vartwo}{(\la\var\vtopptm{\tm}\isubs)\varthree}}}
&\phantom{\hspace{.3cm}} \text{ where } \vtoppctx{\subctx} = (\ppsubctx, \isubs) \text{ and }
\vtopptm{\tmtwo} = \ppsubctxtwop{\varthree} 
\\
\vtopptm{\tm\tmtwo} & 
 \defeq &
\ppsubctxp{\ppsubctxtwop{\vartwo\esub{\vartwo}{\var\varthree}}} 
&\phantom{\hspace{.3cm}}\text{
where } \vtopptm{\tm} =
\ppsubctxp{\var} \text{ and } \vtopptm{\tmtwo} = \ppsubctxtwop{\varthree}, \text{ if } \tm \text{ is not an answer}
\end{array}$
\end{tabular}
}
\caption{The translation from \lvsc to \Lppos.}
\label{fig:translation}	
\end{figure}
%%%
\myparagraph{Subtlety 2: Answers}
It is natural to define $\vtopptm{\cdot}$ as introducing sharing points for every
non-variable sub-term (as in Accattoli et al. \cite{DBLP:conf/lics/AccattoliCC21}), which gives the following definition (where the meta-level substitution $\isub{\var}{\vartwo}$ in the last case is due to the absorption of variables):
\[     \arraycolsep=3pt
\begin{array}{rll@{\hspace{.3cm}}|@{\hspace{.3cm}}rlll}
\vtopptm{\var} & \defeq &\var 
& 
\vtopptm{\tm\tmtwo} & \defeq &
\ppsubctxp{\ppsubctxtwop{\var}\esub{\var}{\vartwo\varthree}} & \text{where }
\vtopptm{\tm} \defeq \ppsubctxp{\vartwo} \text{ and } \vtopptm{\tmtwo} =
\ppsubctxtwop{\varthree} 
\\[4pt]
\vtopptm{\la\var\tm} & \defeq & \vartwo\esub{\vartwo}{\la\var\vtopptm{\tm}} 
&
\vtopptm{\tm\esub{\var}{\tmtwo}} & \defeq &
\ppsubctxp{\vtopptm{\tm}\isub{\var}{\vartwo}} & \text{where }
\vtopptm{\tmtwo} = \ppsubctxp{\vartwo}
\end{array}
\]
Unfortunately, such a definition does not induce a simulation. For instance, consider the following $\egoodsym$-step:
\begin{center}
$\begin{array}{llllllllll} 
\tm &\defeq &(\var\var)\esub{\var}{\la\vartwo\tmtwo} 
&\toegood&
((\la\vartwo\tmtwo)\var)\esub{\var}{\la\vartwo\tmtwo} &=: & \tm'
\end{array}$
\end{center}
Using the above definition of $\vtopptm\cdot$, one would need to have:
\begin{center}
$\begin{array}{lllllllll} 
  \vtopptm\tm&=& \varthree\esub{\varthree}{\var\var}\esub{\var}{\la\vartwo\vtopptm{\tmtwo}}
&   \topposn&
   \varthree\esub{\varthree}{\varfour\var}\esub{\varfour}{\la\vartwo\vtopptm{\tmtwo}}\esub{\var}{\la\vartwo\vtopptm{\tmtwo}}
   &=& \vtopptm{\tm'}
\end{array}$
\end{center}
Note, however, that such a \emph{duplication of ESs} cannot be performed in \Lppos. Therefore, we rather adopt a
modified translation that behaves differently on applied abstractions---more precisely, on applied \emph{answers} (answers are defined in \reffig{weak-vsc-def}, page \pageref{fig:weak-vsc-def}). The new translation is defined in \reffig{translation}. In fact, the translation $\vtopptm\tm$ of terms is defined by mutual induction with the translation $\vtopptm\subctx$ of substitution contexts, which is used in the case of applied answers.
Because of the absorption of variables, the translation $\vtoppctx\sctx$ of substitution contexts $\sctx$ is not simply an evaluation context of \Lppos but a \emph{pair} of an evaluation context $\ppsubctx$ \emph{and} a renaming $\sigma$, that is, a meta-level substitution of variables for variables. For instance, $\vtoppctx{\ctxhole\esub\var{\la\vartwo\tm}\esub\varfour\varthree}= (\ppsubctx, \sigma)$ with $\ppsubctx \defeq \ctxhole\esub\var{\la\vartwo\vtopptm\tm}$ and $\sigma \defeq \isub\varfour\varthree$.

The next lemma shows that such a translation of substitution contexts is compositional. It is proved by a straightforward  induction on $\subctx$.
\begin{lemma}
\label{l:vtopptm-subctx}
Let $\subctxp\tm$ be a \VSC term and $\vtoppctx{\subctx} = (\ppsubctx, \isubs)$. Then  $\vtopptm{\subctxp{\tm}} =
\ppsubctxp{\vtopptm{\tm}\isubs}$.
\end{lemma}

%\myparagraph{Properties of the Translation}

%\begin{lemma}
%\label{l:shape-transl-answers}
%Let $\ans$ be an answer. Then $\vtopptm\ans = \ppsubctxp{\var\esub\var{\la\vartwo\tmtwo}}$ for some $\ppsubctx$ and $\tmtwo$.
%\end{lemma}
%\begin{proof}
%By definition, $\ans = \sctxp{\la\vartwo\tm}$ for some substitution context $\sctx$ and term $\tm$. Note that $\vtopptm\ans = \vtopptm{\la\vartwo\tm} = \var\esub\var{\la\vartwo\vtopptm\tm}$. Then apply \reflemma{vtopptm-subctx}.
%\end{proof}

\myparagraphmed{Simulation} Core reduction $\tocore$ is made out of three kinds of steps, namely $\tomo$, $\toegood$, and $\toevar$. Given the special role of answers in the definition of $\vtopptm\cdot$, the proof of the simulation becomes tricky when core steps can turn an applied non-answer into an applied answer. This can happen with $\tomo$ and $\toegood$ steps, which are then discussed in detail in the next paragraphs. Rule $\toevar$, instead, does not alter whether sub-terms are answers, and so the proof that $\toevar$ steps are absorbed is smooth.

\begin{toappendix}
\begin{lemma}[Absorption of variable exponentials\NoteProof{l:evar-absorption}]
\label{l:evar-absorption}
Let $\tm$ and $\tmtwo$ be \VSC terms. If $\tm \toevar \tmtwo$ then $\vtopptm{\tm} = \vtopptm{\tmtwo}$.
\end{lemma}
\end{toappendix}

\myparagraphmed{Subtlety 3: Simulation of Multiplicative Steps} Root multiplicative steps are simulated smoothly.

\begin{toappendix}
\begin{lemma}[Simulation of root multiplicative steps\NoteProof{l:root-mult-simulation}]
\label{l:root-mult-simulation}
Let $\tm$ and $\tmtwo$ be \VSC terms. If $\tm \rtom \tmtwo$ then $\vtopptm{\tm} \toposm \vtopptm{\tmtwo}$.
\end{lemma}
\end{toappendix}

A complication arises for the contextual closure of multiplicative steps, because in a root step $\subctxp{\la\var\tm}\tmtwo \rtom \subctxp{\tm\esub\var\tmtwo}$ the redex is not an answer but the reduct might be one, if $\tm$ is an abstraction. Thus, if the root step is applied to a further argument $\tmthree$, the reduction turns an applied non-answer into an applied answer, changing the clause of the translation that is used for the application to $\tmthree$. This phenomenon is handled by doing two additional rewriting steps in \Lppos. The simplest case is the following one, where $\tm=\vartwo$, $\tmtwo = \varthree$, $\tmthree=\varfour$, and $\subctx = \ctxhole$ and $\var'$, $\vartwo'$, $\varthree'$ are variables introduced by the translation:
\begin{center}
\begin{tikzpicture}[ocenter]
		\node at (0,0)[align = center](source){\normalsize $(\la\var\la\vartwo\vartwo)\varthree\varfour$};
		\node at (source.center)[right = 345pt](source-right){\normalsize $(\la\vartwo\vartwo)\esub\var\varthree \varfour$};
		\node at (source.center)[below = 25pt](source-down){\normalsize $\var'\esub{\var'}{\vartwo'\varfour}\esub{\vartwo'}{(\la\var \varthree' \esub{\varthree'}{\la\vartwo\vartwo})\varthree}$};
		\node at (source-right|-source-down)(target){\normalsize $\var'\esub{\var'}{(\la\vartwo\vartwo)\varfour}$};
		\node at \med{source-down.east}{target.west}[anchor=center](midnode){};
		\node at (midnode.center)[left = 16pt](midnodeLeft){\normalsize 
		$\var'\esub{\var'}{\varthree'\varfour} \esub{\varthree'}{\la\vartwo\vartwo}$};
		\node at (midnode.center)[right = -4pt](midnodeRight){\normalsize $\var'\esub{\var'}{(\la\vartwo\vartwo)\varfour} \esub{\varthree'}{\la\vartwo\vartwo}$};
		
		\draw[->](source) to node[above] {\scriptsize $\osym\msym$} (source-right);
		
		\draw[->, dotted](source) to node[left] {\scriptsize $\vtopptm\cdot$}(source-down);
		\draw[->, dotted](source-right) to node[right] {\scriptsize $\vtopptm\cdot$}(target);
		
		\draw[->](source-down) to node[above] {\scriptsize $\osym\mpsym$} (midnodeLeft);
		\draw[->](midnodeLeft) to node[above] {\scriptsize $\osym\epsym$} (midnodeRight);
		\draw[->](midnodeRight) to node[above] {\scriptsize $\osym\gcpsym$} (target);
%		\draw[<->](tarcal) to (tartam);
	\end{tikzpicture}
\end{center}

In general, we have the following simulation of multiplicative steps, where the first case isolates exactly when applied non-answers are turned into applied answers. 
\begin{toappendix}
\begin{proposition}[Simulation of $\tomo$ steps\NoteProof{prop:sim-mult-steps}]
\label{prop:sim-mult-steps}
  Let $\tm$ and $\tmtwo$ be \VSC terms and $\tm \rtom \tmtwo$.
\begin{enumerate}
  \item If $\tmtwo$ is an answer and $\isgood{\octx}$ then $\vtopptm{\octxp{\tm}} \tomp \toep\togcp
\vtopptm{\octxp{\tmtwo}}$;
  \item Otherwise, $\vtopptm{\octxp{\tm}} \tomp \vtopptm{\octxp{\tmtwo}}$.
\end{enumerate}
\end{proposition}
\end{toappendix}

%%%
\begin{figure}
\centering
\fbox{
\begin{tabular}{cc}
$\begin{array}{r r@{\ }l@{\ }l}
\textsc{\Good exp. root rule 1} &\phantom{\hspace{.3cm}}
\goodctxfp{\var}\esub\var{\subctxp{\la\vartwo\tm}} &
\rtoegooda &
\subctxp{\goodctxfp{\la\vartwo\tm}\esub{\var}{\la\vartwo\tm}}
\\
\textsc{\Good exp. root rule 2} &\phantom{\hspace{.3cm}}
\subctxfirstp{\subctxsecfp{\var}\esub{\var}{\subctxthirdp{\la\vartwo\tm}}}\tmtwo &
\rtoegoodb &
\subctxfirstp{\subctxthirdp{\subctxsecfp{\la\vartwo\tm}\esub{\var}{\la\vartwo\tm}}}\tmtwo
\end{array}$
			\\[15pt]
			\hline
			\\[-12pt]
					$\begin{array}{rl}
					\textsc{Ctx closure}
					&\phantom{\hspace{.3cm}}
					\RightLabel{$\asym{\in}\{\egoodasym,\egoodbsym\}$}
					\AxiomC{$\tm \rootRew{\asym} \tm'$}
					\UnaryInfC{$\octxp\tm \Rew{\wsym\asym} \octxp{\tm'}$}
					\DisplayProof
					\end{array}$
\end{tabular}
}
\caption{Two root rules for $\toegood$.}
\label{fig:dissecting-egood}
\end{figure}
%%%
\myparagraphmed{Subtlety 4: Simulation of Useful Exponential Steps} Exponential steps  can turn applied non-answers into applied answers too: these cases actually are the very definition of useful exponential steps. In contrast to the multiplicative case, the simulation does not need extra steps, it simply maps one $\toegood$ step in \lvsc to one $\topose$ step in \Lppos. What is tricky in this case is that the definition of useful steps needs the surrounding open context of the root step, as explained in \refsect{dissecting}, so it does not seem possible to prove the simulation for a root case and then extending the property with an induction on the context closure.

To get around this issue, in \reffig{dissecting-egood} we give an alternative definition of $\toegood$ resting on \emph{two} root rules, where the second rule captures the cases when the argument of the created redex is provided by the context. With these two root rules, useful exponential steps can then be defined via a closure by \emph{any} open context, thus bypassing the global aspect of the definition that we gave in \refsect{dissecting}. The reader probably wonders why we did not do it already in \refsect{dissecting}. The reason is that the global definition is preferable for proving local postponement (\refprop{local-postponement-bad}). The alternative definition is justified by the following lemma.

\begin{toappendix}
\begin{lemma}[Alternative presentation of useful steps\NoteProof{l:egood-split}]
\label{l:egood-split}
$\toegood = \toegooda \cup \toegoodb$.
\end{lemma}
\end{toappendix}

The simulation is then proved smoothly via the alternative definition.
\begin{toappendix}
\begin{proposition}[Simulation of useful exponential steps\NoteProof{prop:sim-us-exp-steps}]
Let $\tm$ and $\tmtwo$ be \VSC terms. If $\tm \toegooda \tmtwo$ or $\tm \toegoodb \tmtwo$ then $\vtopptm{\tm} \topose
\vtopptm{\tmtwo}$.\label{prop:sim-us-exp-steps}
\end{proposition}
\end{toappendix}

\myparagraphsmall{Summing Up} We can now put together the simulations of single core steps, and also iterate over reduction sequences. Since at times one source multiplicative step is simulated by more than one target step, the simulation does not preserve evaluation lengths.  An important point to note, however, is that the number of multiplicative steps---which is the cost model of \lvsc---is preserved. More generally, the increment in length is only linear.
\begin{theorem}[Simulation of core sequences]
\label{thm:vsc-to-lppos}
  Let $\deriv: \tm \togoodvscn \tmp$ be a reduction sequence in \lvsc.
  Then there exists \ignore{a reduction sequence} $\derivtwo: \vtopptm{\tm} \topposn
\vtopptm{\tmp}$ in \Lppos such that $\sizeposm{\derivtwo} = \sizemo{\deriv}$ and $\sizep\deriv{\osym\msym,\osym\egoodsym}\leq \size\derivtwo \leq 3\cdot\size\deriv$.
\end{theorem}

%%%%%%%%%%%%%%%%%%%%%%
%\input{08-core_normal_forms}
%%%%%%%%%%%%%%%%%%%%%%
\section{Core Normal Forms and Termination Equivalence} 
\label{sect:termination-equiv}
In this section, we show that the translation $\vtopptm\cdot$ preserves and reflects termination. Reflection is a consequence of the simulation theorem: if $\vtopptm\tm$ terminates then $\tm$ cannot diverge, because $\vtopptm\tm$ can simulate it. Preservation instead is proved by showing that $\tocore$ normal forms are mapped to (non-erasing) positive normal forms, \ie, $\toepngc$-normal forms, which is proved via a characterization of core normal forms.

\myparagraphmed{Characterization of Core Normal Forms} For the technical characterization of core normal forms, we need a few auxiliary definitions. We start by defining two sets of variables for terms. 

\begin{definition}
The set $\ofv\tm$ of \emph{open free variables} of a VSC term $\tm$ is the set of variables of $\tm$ having occurrences out of all abstractions, formally defined as follows:
\[\begin{array}{rll rlll}
\ofv{\var} &\defeq& \set{\var} 
& \phantom{\hspace{.3cm}}\phantom{\hspace{.3cm}}
\ofv{\tm\tmtwo} &\defeq& \ofv{\tm} \cup \ofv{\tmtwo}
\\
\ofv{\la\var\tm} &\defeq& \emptyset 
&\phantom{\hspace{.3cm}}\phantom{\hspace{.3cm}}
\ofv{\tm\esub\var\tmtwo} &\defeq& (\ofv{\tm}\setminus \set\var) \cup \ofv{\tmtwo}
\end{array}
\]
The set $\aofv\tm$ of \emph{applied open free variables} of a VSC term $\tm$ is the set of variables of $\tm$ having applied occurrences out of all abstractions, formally defined as follows:
\[\begin{array}{rll | rlll}
\aofv\var = \aofv{\la\var\tm} &\defeq& \emptyset \phantom{\hspace{.3cm}}
&\phantom{\hspace{.3cm}}
\aofv{\tm\tmtwo} &\phantom{\hspace{.3cm}} \defeq & \aofv\tm \cup \aofv\tmtwo \cup \set{\var} & \text{ if } \tm = \sctx\ctxholefp\var 
\\
\aofv{\tm\esub\var\tmtwo} & \defeq & (\aofv\tm \setminus \set\var) \cup \aofv\tmtwo\phantom{\hspace{.3cm}}
& \phantom{\hspace{.3cm}}
\aofv{\tm\tmtwo} &\phantom{\hspace{.3cm}} \defeq & \aofv\tm \cup \aofv\tmtwo & \text{ otherwise }
\end{array}\]
\end{definition}

We also need a weakened notion of answer.
\begin{definition}
An \emph{\aanswer} is an answer or a \VSC term of the form
$\sctxp{\sctxtwop\var\esub\var{\tm}}$ where $\tm$ is an answer.
\end{definition}
Finally, we can provide a characterization of core normal forms, based on the following grammar.
\begin{center}
$\begin{array}{lllllllll}
\multicolumn{4}{c}{\textsc{Grammar of core normal terms}}
\\
\ntm &= &\val 
\\
&&\mid \ntm \ntmtwo &\mbox{with }\ntm\mbox{ not an \aanswer}
\\
&&\mid \ntm \esub\var{\ntmtwo} &\mbox{with }\ntmtwo = \sctxp{\la\vartwo\tm}\mbox{ and }\var\notin\aofv\ntm
\\
&&\mid \ntm \esub\var{\ntmtwo} &\mbox{with }\ntmtwo = \sctxp{\vartwo}\mbox{ and }\var\notin\ofv\ntm
\\
&&\mid \ntm \esub\var{\ntmtwo} &\mbox{with }\ntmtwo = \sctxp{\tm\tmtwo}
\end{array}$
\end{center}

\begin{toappendix}
\begin{proposition}[Characterization of core normal forms\NoteProof{prop:charac-core-nfs}]
\label{prop:charac-core-nfs}
Let $\tm$ be a VSC term. $\tm$ is $\tocore$-normal if and only if it is a $\ntm$ term.
\end{proposition}
\end{toappendix}
Via a few technical lemmas \withproofs{in \refapp{app-translation}}\withoutproofs{(in the tech report \cite{accattoli:hal-04606194})}, we obtain the following preservation property.
\begin{toappendix}
\begin{proposition}[Preservation of core normal forms\NoteProof{prop:preserv-core-nfs}]
\label{prop:preserv-core-nfs}
Let $\tm$ be a \VSC term. If $\tm$ is $\tocore$-normal then $\vtopptm{\tm}$ is $\toepngc$-normal.
\end{proposition}
\end{toappendix}

\begin{toappendix}
\begin{theorem}[Termination equivalence of Core \lvsc and \Loxposnospace\NoteProof{prop:preserv-core-nfs}]
\label{thm:core-term-equiv}
Let $\tm$ be a \VSC term. 
\begin{enumerate}
\item $\tm$ has a diverging $\tocore$ sequence if and only if $\vtopptm{\tm}$ has a diverging $\toepos$ sequence.
\item $\tm$ is $\tocore$-weakly normalizing if and only if $\vtopptm{\tm}$ is $\toepos$-weakly normalizing.
\end{enumerate}
\end{theorem}
\end{toappendix}
As a corollary, we can prove \emph{uniform normalization} (see \refsect{rewriting-notions}) for \lvsc and Core \lvsc. Uniform normalization follows immediately from the diamond property, thus it holds for \Lppos (\refthm{lppos-diamond}). Concerning \lvsc and Core \lvsc, the proof is instead not immediate, because they are not diamond (see the diagram at page \pageref{lvsc-not-diamond}). But we can obtaining it by lifting the one of \Lppos via the proved termination equivalences.
\withproofs{\begin{toappendix}
\begin{corollary}[\NoteProofNoComma{coro:unif-norm}]
\label{coro:unif-norm}
\lvsc and Core \lvsc are uniformly normalizing.
\end{corollary}
\end{toappendix}}
\withoutproofs{
\begin{corollary}
\label{coro:unif-norm}
\lvsc and Core \lvsc are uniformly normalizing.
\end{corollary}
}

%\myparagraphsmall{Complements}
%It is not needed for the simulation of the previous section, but it is easy to show that the translation behaves well with respect to garbage collection. The collection of abstractions is simulated in one step, and the collection of variables is absorbed. The two properties are proved smoothly, since GC cannot alter whether sub-terms are answers.
%
%\begin{toappendix}
%\begin{proposition}[Simulation and Absorption of GC\NoteProof{prop:GC-simulation-and-absorption}]
%Let $\tm$ and $\tmtwo$ be \VSC terms.\label{prop:GC-simulation-and-absorption}
%\begin{enumerate}
%  \item \emph{GC simulation}: if $\tm \togcabs \tmtwo$ then $\vtopptm{\tm} \togctwo \vtopptm{\tmtwo}$.
%  \item \emph{GC absorption}: if $\tm \togcvar \tmtwo$ then $\vtopptm{\tm} = \vtopptm{\tmtwo}$.
%\end{enumerate}
%\end{proposition}
%\end{toappendix}

%%%%%%%%%%%%%%%%%%%%%%
%\input{09-conclusions}
%%%%%%%%%%%%%%%%%%%%%%
\section{Conclusions}
\label{sect:conclusions}
This paper studies Wu's positive $\l$-calculus \Lpos, which at first sight looks simply as yet another call-by-value $\l$-calculus with sharing. It has, however, a new feature that distinguishes it among similar calculi, called here \emph{compactness} and concerning the treatment of variables.

Our main contribution is showing that compactness allows one to elegantly capture the essence of useful sharing (in an open setting), circumventing the main technicalities of this notion. What is remarkable is that \Lpos has not arisen as an incremental refinement of useful sharing, but as an outcome of the completely unrelated study of focalization for minimal intuitionistic logic by Miller and Wu \cite{DBLP:conf/csl/0001W23}. 

We believe that the positive $\l$-calculus is a sharp tool deserving to be studied further, in particular with respect to program transformations and optimizations, and also endowed with call-by-need evaluation.

An aspect that we have left for future work is the efficient implementation of the meta-level renamings involved in the multiplicative rewriting rule, which can be computationally costly if done without care. We expect it to be doable efficiently via an appropriate abstract machine.

%%%
%\newpage
\bibliographystyle{./entics}
\bibliography{main.bbl}
%\section{An appendix}
%Any appendices should be included after the references, as is done here. The appendix starts with the command \verb+\appendix+, after which sections can be included; these are lettered, rather than numbered. 
%\section*{Another appendix}
%One can also use \verb+\section*{...}+ to create an appendix without a letter attached. 
\end{document}